\documentclass[twocolumn]{aastex631} 
\usepackage{multirow}
\usepackage{hyperref}
\usepackage{graphicx}
\shorttitle{Sausage cluster at sub-GHz frequencies}
\shortauthors{Raja et al.}

\begin{document}

\title{uGMRT sub-GHz view of the Sausage cluster diffuse radio sources}

\correspondingauthor{Ramij Raja}
\email{ramij.amu48@gmail.com}

\author[0000-0001-8721-9897]{Ramij Raja}
\affiliation{Centre for Radio Astronomy Techniques and Technologies, Department of Physics and Electronics,\\ Rhodes University, Makhanda 6140, South Africa}

\author[0000-0003-1680-7936]{Oleg M. Smirnov}
\affiliation{Centre for Radio Astronomy Techniques and Technologies, Department of Physics and Electronics,\\ Rhodes University, Makhanda 6140, South Africa}
\affiliation{South African Radio Astronomy Observatory, 2 Fir Street, Black River Park, Observatory, Cape Town 7925, South Africa}
\affiliation{Institute for Radioastronomy, National Institute of Astrophysics (INAF IRA), Via Gobetti 101, 40129 Bologna, Italy}

\author[0000-0002-8476-6307]{Tiziana Venturi}
\affiliation{Institute for Radioastronomy, National Institute of Astrophysics (INAF IRA), Via Gobetti 101, 40129 Bologna, Italy}
\affiliation{Centre for Radio Astronomy Techniques and Technologies, Department of Physics and Electronics,\\ Rhodes University, Makhanda 6140, South Africa}

\author[0000-0002-1372-6017]{Majidul Rahaman}
\affiliation{Institute of Astronomy, National Tsing Hua University, Hsinchu 300013, Taiwan}

\author[0000-0003-3269-4660]{H.-Y. Karen Yang}
\affiliation{Institute of Astronomy, National Tsing Hua University, Hsinchu 300013, Taiwan}
\affiliation{Physics Division, National Center for Theoretical Sciences, Taipei 106017, Taiwan}



\begin{abstract}
CIZA J2242.8+5301, or the Sausage cluster, is well studied over a range of frequencies. Since its first discovery, a lot of interesting features and unique characteristics have been uncovered. In this work, we report some more new morphological features using the uGMRT band-3 and band-4 data. 
In the north relic, we observe variation in spectral index profiles across the relic width from the east to west, which may indicate a decrease in downstream cooling rate in that direction. We re-confirm the presence of an additional $\sim 930$ kpc relic in the north. We classify the filamentary source in the downstream region to be a narrow angle tail (NAT) radio galaxy. 
The bright arc in the east relic shows sub-structure in the spectral index profile, which may indicate the presence of finer filaments. We further report the presence of a double-strand structure in the east relic similar to the `Toothbrush’ relic.
We categorize the bright `L’ shaped structure in the southern relic to be a NAT radio galaxy, as well as trace the actual $\sim 1.1$ Mpc relic component. We re-confirm the existence of the faint southern extent, measuring the relic length to be $\sim 1.8$ Mpc. Furthermore, we suggest the southern relic to be a union of individual component relics rather than a single giant filamentary relic. Lastly, based on the morphological symmetry between northern and southern relics, we suggest a schematic shock structure associated with the merger event in an attempt to explain their formation scenario. 

\end{abstract}



\section{Introduction} \label{sec:intro}
In the hierarchical structure formation process of the Universe, smaller structures like galaxy groups and sub-clusters are accreted via the cosmic filaments, driving the growth of galaxy clusters located at the cross-section of these cosmic filaments \citep[e.g.,][]{Press1974ApJ...187..425P,Springel2006Natur.440.1137S}. 
These merging events release massive amounts of energy ($\sim 10^{64}$ erg) into the intracluster medium (ICM) in the form of large-scale shocks and turbulence \citep{Sarazin2002ASSL..272....1S}.
Most of this energy heats up the ICM to $10^7 - 10^8$ K, as observed in X-ray observations \citep[e.g.,][]{Rahaman2022MNRAS.509.5821R}, while a fraction of this energy is expended in magnetic field amplification and (re)acceleration of particles resulting in the formation of \textit{halos} and \textit{relics} detected in radio observations \citep[see][for theoretical and observational review]{Brunetti2014IJMPD..2330007B,vanWeeren2019SSRv..215...16V,Paul2023JApA...44...38P}.

\textit{Radio halos} are Mpc-scale centrally located diffuse radio emissions that generally follow the X-ray brightness distribution of the thermal ICM component \citep[e.g.,][]{Venturi2013A&A...551A..24V,Shimwell2014MNRAS.440.2901S}. They are low-surface brightness, steep spectrum ($\alpha < -1$, with $S_{\nu} \propto \nu^{\alpha}$) sources predominantly observed in merging cluster \citep[e.g.,][]{DiGennaro2021A&A...654A.166D}. The currently favored formation mechanism of radio halos is considered to be re-acceleration of \textit{in situ} relativistic electrons via merger-induced turbulence \citep[e.g.,][]{Brunetti2001MNRAS.320..365B,Petrosian2001ApJ...557..560P,Brunetti2007MNRAS.378..245B,Brunetti2016MNRAS.458.2584B}.

\textit{Radio relics} are $\sim 1-2$ Mpc scale elongated diffuse radio objects located at the cluster periphery \citep[e.g.,][]{vanWeeren2016ApJ...818..204V,DiGennaro2018ApJ...865...24D,Rajpurohit2021A&A...646A.135R}. It is observed that they trace shock waves in the ICM, generated during major merger events \citep[e.g.,][]{Finoguenov2010ApJ...715.1143F,Akamatsu2015A&A...582A..87A,Andrade-Santos2019ApJ...887...31A}. The observed radio emission has steep spectra  \citep[$-1\lesssim \alpha \lesssim-1.5$; e.g.,][]{vanWeeren2012A&A...546A.124V,deGasperin2015MNRAS.453.3483D} and is highly polarized at GHz frequencies \citep[$\sim20\%-50\%$; e.g.,][]{Bonafede2009A&A...494..429B,vanWeeren2010Sci...330..347V,Rajpurohit2022A&A...657A...2R}. 
So far, the most likely mechanism behind radio relics formation is considered to be the diffusive shock acceleration \citep[DSA; e.g.,][]{Ensslin1998A&A...332..395E,Blasi1999APh....12..169B,Dolag2000A&A...362..151D,Caprioli2012JCAP...07..038C}, which reproduces the observed power-law radio spectra \citep[e.g.,][]{Hoeft2007MNRAS.375...77H}. However, there are examples of radio relic spectra that deviate from the simple power-law model and show steepening at high-frequencies \citep[e.g.,][]{Malu2016Ap&SS.361..255M,Stroe2016MNRAS.455.2402S} and flattening ($\alpha > -1$) at low-frequencies \citep[e.g.,][]{Kale2010ApJ...718..939K,vanWeeren2012A&A...543A..43V,Trasatti2015A&A...575A..45T}{}{}, although later studies have refuted some of these claims, reporting single power-law fit even up to much higher frequencies \citep[e.g.,][]{Loi2017MNRAS.472.3605L,Loi2020MNRAS.498.1628L,Rajpurohit2020A&A...642L..13R,Rajpurohit2022ApJ...927...80R}{}{}. 
Other mechanisms proposed for relic formation are diffusive shock re-acceleration \citep[e.g.,][]{Shimwell2015MNRAS.449.1486S,vanWeeren2017NatAs...1E...5V,Botteon2020A&A...634A..64B}, multiple shock acceleration \citep[e.g.,][]{Inchingolo2022MNRAS.509.1160I,Smolinski2023MNRAS.526.4234S}. 

In this paper, we have investigated multiple diffuse radio objects in the Sausage cluster at two sub-GHz frequencies in an attempt to re-validate the previous results as well as discover new and interesting characteristics. 
In the following, we briefly outline the previous works on this cluster in Sect. \ref{sec:sausage}. 
A brief description of the observations and data analysis method is presented in Sect. \ref{sec:obs}. 
The various diffuse radio objects in the northern relic region are discussed in detail in Sect. \ref{sec:NR}.
A similar detailed analysis of the southern relic is presented in Sect. \ref{sec:SR}. 
Finally, a summary of the results and concluding remarks are presented in Sect. \ref{sec:conclude}.

Throughout this paper, we adopt a $\Lambda$CDM cosmology with $H_0 = 70$ km s$^{-1}$ Mpc$^{-1}$, $\Omega_{\mathrm{m}} = 0.3$ and $\Omega_\Lambda = 0.7$. At the cluster redshift $z = 0.1921$ \citep{Kocevski2007ApJ...662..224K}, $1\arcsec$ corresponds to a physical scale of $3.197$ kpc.

\section{The Sausage cluster} \label{sec:sausage}
CIZA J2242.8+5301 (hereafter CIZA2242) or the `Sausage cluster' is a massive cluster with a signature merger state located at redshift $z=0.1921$, first observed with \textit{ROSAT} in X-ray \citep[][]{Kocevski2007ApJ...662..224K}. The X-ray luminosity of the cluster within $R_\mathrm{500} = 1.2$ Mpc is $L_\mathrm{[0.1-2.4 keV]}^\mathrm{500} = (7.7 \pm 0.1) \times 10^{44}$ erg s$^{-1}$ \citep{Hoang2017MNRAS.471.1107H}. 
Since its discovery, a lot of studies has been performed across the electromagnetic spectrum ranging from radio \citep[e.g.,][]{vanWeeren2010Sci...330..347V,Stroe2016MNRAS.455.2402S,Hoang2017MNRAS.471.1107H,DiGennaro2018ApJ...865...24D,DiGennaro2021ApJ...911....3D}{}{} and optical \citep[e.g.,][]{Dawson2015ApJ...805..143D,Jee2015ApJ...802...46J,Okabe2015PASJ...67..114O,Golovich2019ApJ...882...69G}{}{} to X-ray \citep[e.g.,][]{Ogrean2013MNRAS.429.2617O,Ogrean2014MNRAS.440.3416O,Akamatsu2015A&A...582A..87A}{}{}. Furthermore, because of the presence of the spectacular `Sausage' relic, this cluster has been a test example for various cluster shock physics simulations \citep[e.g.,][]{vanWeeren2011MNRAS.418..230V,Kang2015ApJ...809..186K,Donnert2016MNRAS.462.2014D,Boss2023ApJ...957L..16B}{}{}.

The presence of diffuse radio emission in this cluster was first reported by \citet{vanWeeren2010Sci...330..347V}. With the GMRT, WSRT and VLA observations, they detected both north and south relics at an excellent resolution, located $\sim 1.5$ Mpc north and south of the cluster center. They further reported that the $\sim2$ Mpc scale north relic is of extremely narrow width of $\sim55$ kpc with spectral gradient of $-0.6$ to $-2$ (within $0.61$ to $2.3$ GHz) across the relic width. The integrated relic spectrum was reported to be a single power law with $\alpha_\mathrm{610 MHz}^\mathrm{2.3 GHz} = -1.08 \pm 0.05$, and the emission is strongly polarized \citep[$50-60\%$;][]{vanWeeren2010Sci...330..347V,Kierdorf2017A&A...600A..18K}. They also reported the presence of a faint radio halo spanning $\sim3.1$ Mpc between the two radio relics, which was also later confirmed by \citet{Stroe2013A&A...555A.110S,Hoang2017MNRAS.471.1107H,DiGennaro2018ApJ...865...24D}. Subsequently, \citet{Stroe2013A&A...555A.110S} reported spectral study of four radio relics in the cluster using GMRT and WSRT observations, discovering the presence of steep spectrum in all of them.  
Later \citet{Hoang2017MNRAS.471.1107H} resolved the discrepancy between X-ray and radio derived Mach numbers of the relics. Furthermore, with the \textit{Suzaku} X-ray observations, they detected a temperature jump at the Eastern relic position, suggesting a Mach $\mathcal{M} \sim 2.4$ shock associated with the Eastern relic. Subsequently, \citet{DiGennaro2018ApJ...865...24D} using high resolution VLA observations, discovered complex morphology of the diffuse radio sources along with filamentary structures in the `Sausage' relic.

\section{Observations and Data Analysis} \label{sec:obs}

\subsection{uGMRT band-3 data reduction} \label{subsec:band3_data}
The uGMRT band-3 data used in this study (project code: ddtC231) were observed during July-September 2022. The observations were made with $\sim200$ MHz bandwidth divided into 4096 channels for a total of $\sim12.1$ hours of on-source time. 

The uGMRT band-3 data reduction was done using \texttt{SPAM} \citep[Source Peeling and Atmospheric Modeling; ][]{Intema2009A&A...501.1185I,Intema2017A&A...598A..78I}, a Python-based pipeline employing \texttt{AIPS} \citep{Greisen2003ASSL..285..109G} for calibration and imaging. It follows the standard radio interferometric data reduction steps, performing RFI (Radio Frequency Interference) flagging, bandpass, and gain calibration. Subsequently, it proceeds with a few rounds of self-calibration followed by direction-dependent calibration \citep[for more details, see][]{Intema2017A&A...598A..78I}. 

The observations in band-3 were performed over five different nights. Therefore, the first phase of the above data reduction procedure (known as pre-calibration) was performed separately for individual observing sessions. More specifically, each observing session data was first split into six $\sim33.3$ MHz chunks with the \texttt{SPAM} task \textit{split\_wideband\_uvdata} followed by basic 1GC calibration i.e., flagging, bandpass, and gain calibration performed with the task \textit{pre\_calibrate\_wideband\_targets}. After that, we processed the same subbands from five observing sessions together in the main pipeline with the task \textit{process\_wideband\_target}. In the output, we got calibrated \textit{uv-}data and images corresponding to six subbands combined over five observing sessions. We checked these output images and found one of the subband data to be of low quality compared to the rest. Therefore, we proceeded with the rest of the five subbands for our imagining and analysis purposes. The output \textit{uv-}data in FITS format are converted into MS format with \texttt{CASA} \citep[Common Astronomy Software Applications;][]{McMullin2007ASPC..376..127M,CASA2022PASP..134k4501C}. The details of the imaging process are discussed in Sect. \ref{subsec:imaging}.

\subsection{uGMRT band-4 data reduction} \label{subsec:band4_data}
The uGMRT band-4 data used in this study (project code: 36\_006) was observed on 1 September 2019. The observation was made with $\sim400$ MHz bandwidth divided into 4096 channels for a total of $\sim6.7$ hours of on-source time.  

For uGMRT band-4 data reduction, we followed the same procedure as described above for band-3 data. Here, there is a single observing session which is split into eight $50$ MHz frequency chunks. Individual subbands are processed separately, starting with the pre-calibration step. At this stage, two subbands had bad data, resulting in failed calibration. Therefore, we proceeded with the remaining 6 subbands and got the final calibrated \textit{uv-}data along with subband images. Upon further inspection of these images, we further discarded another subband due to low data quality. Finally, we proceeded with 5 subbands for combined imaging for this work, which is discussed in Sect. \ref{subsec:imaging}.

\begin{figure*}
    \centering
    \includegraphics[width=2.1\columnwidth]{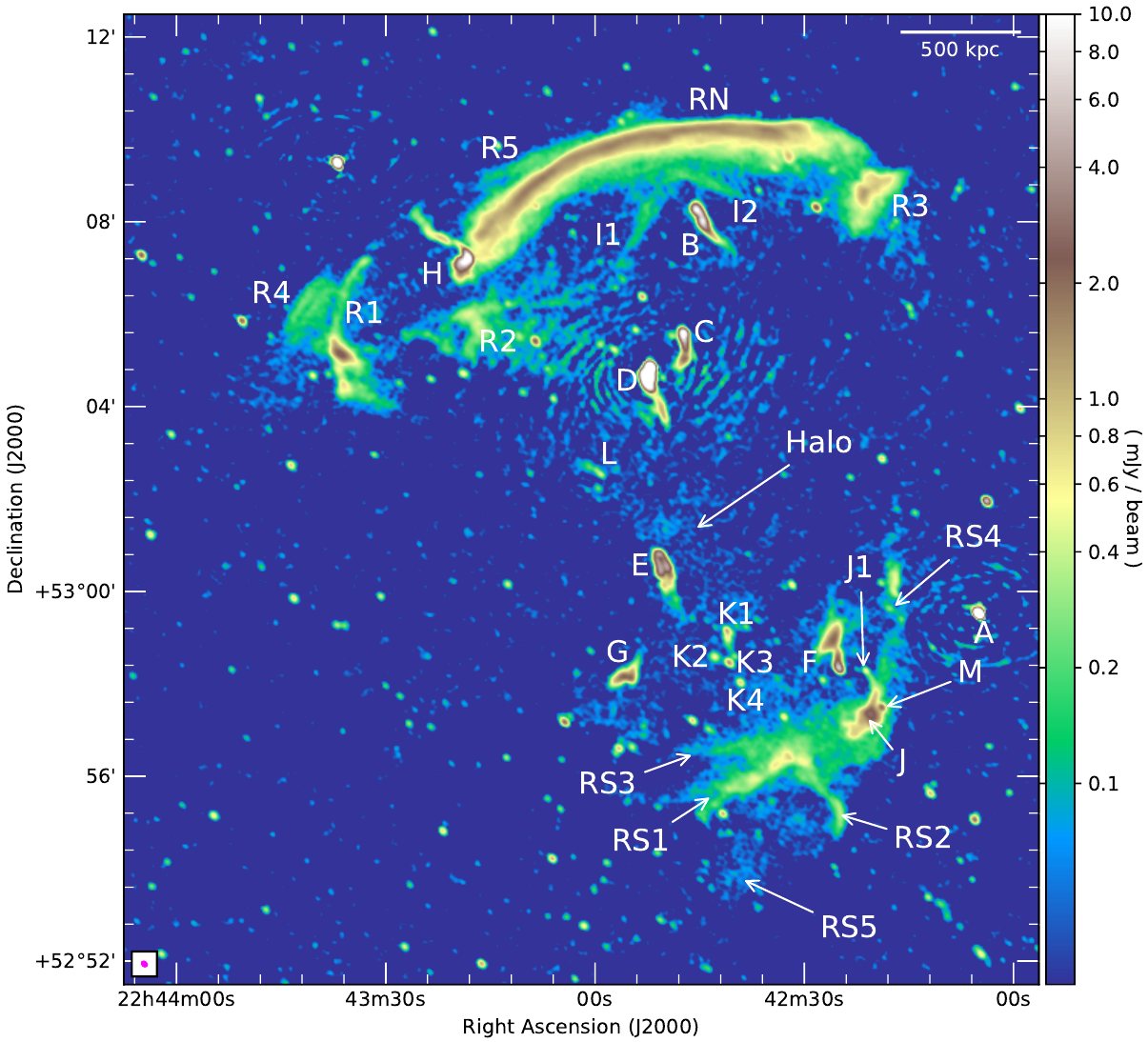}
    \caption{The 400 MHz high-resolution uGMRT image of the Sausage cluster. The image properties are listed in Table \ref{tab:image_prop}, see label IM1. The source labels are adopted from \citet{DiGennaro2018ApJ...865...24D}.}
    \label{fig:Labels}
\end{figure*}

\begin{figure*}[]
    \centering
    \begin{tabular}{cc}
    \includegraphics[width=\columnwidth]{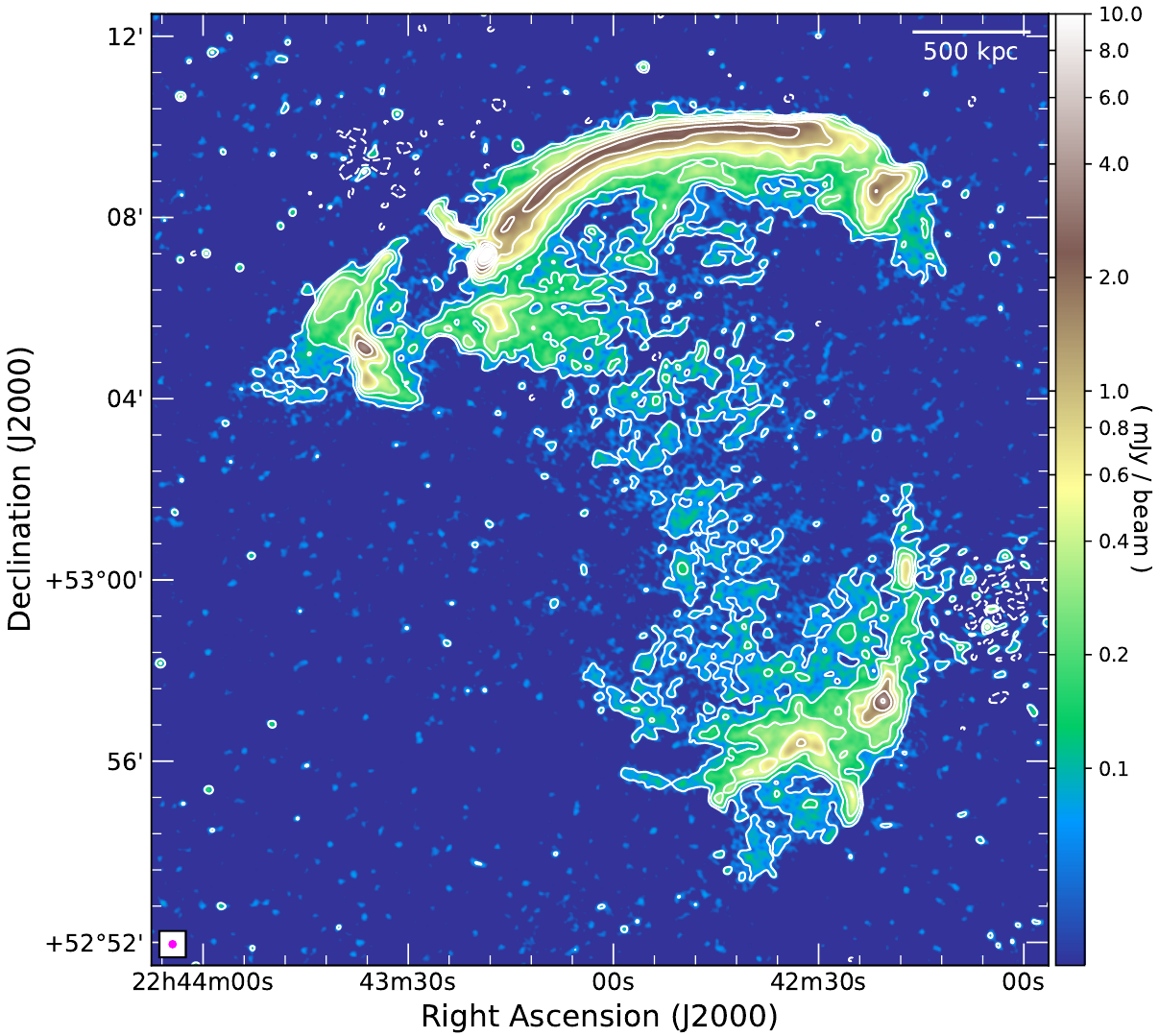} & 
    \includegraphics[width=\columnwidth]{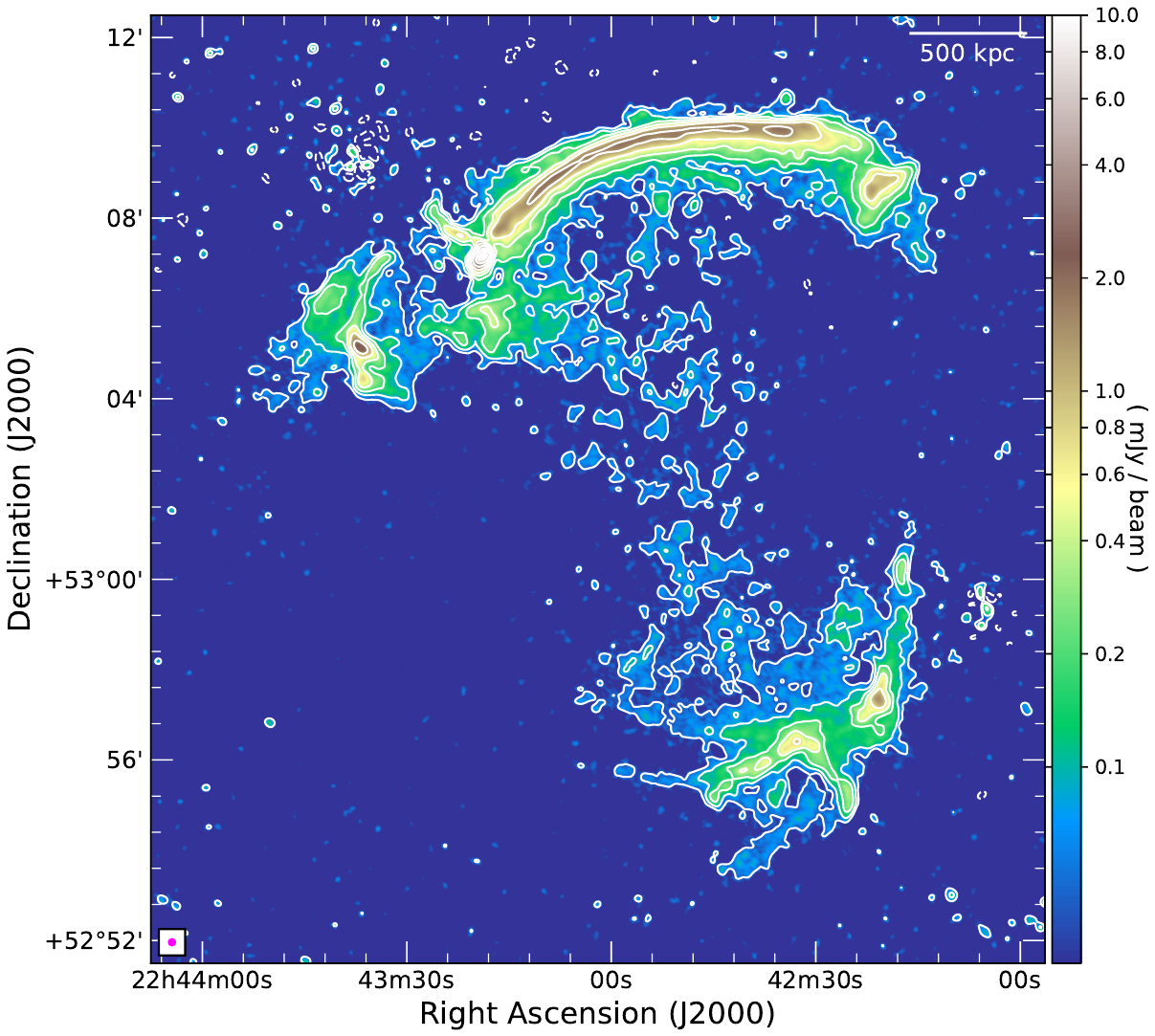}
    \end{tabular}
    \caption{High-resolution ($8.5\arcsec$) uGMRT images at 400 (\textit{left}) and 675 MHz (\textit{right}) overlaid with contours. The image properties are listed in Table \ref{tab:image_prop}, see label IM2 and IM5. The contours are drawn at levels $[-1, 1, 2, 4, 8,...] \times 3\sigma_{\rm rms}$. Negative contours are shown with dashed line.}
    \label{fig:400_700_8p5}
\end{figure*}

\subsection{Imaging} \label{subsec:imaging}
Imaging of the calibrated band-3 and band-4 \textit{uv-}data were done using the \texttt{WSCLEAN} \citep{offringa-wsclean-2014,Offringa2017MNRAS.471..301O} package. The imaging strategy for both these data sets is similar. We produced full-resolution images exploring the spectral fitting parameter \texttt{nterms = 1 - 4} and found images with \texttt{nterms = 4} to be of slightly better quality than the rest. We also experimented with different Briggs \texttt{robust} parameters and finally chose \texttt{robust = 0} as the best compromise between sensitivity and resolution. 

There are a lot of radio galaxies (RGs) and compact radio sources (CRS) embedded within the cluster diffuse objects (Fig. \ref{fig:Labels}). Therefore, to remove the contamination of these sources, we subtracted them from the \textit{uv-}data by properly modeling their clean components. For this purpose, we excluded smaller baselines ($\lesssim1$k$\lambda$) and chose \texttt{robust = -1} at the time of imaging to reduce the diffuse emission contribution to the modeling. We also provided custom masks for better modeling of the clean components, which were generated using \texttt{BREIZORRO}\footnote{\url{https://github.com/ratt-ru/breizorro}}. 

After subtracting the RGs and CRS from the \textit{uv-}data, we produced images of both high and low resolution. 
The high-resolution images were restored at the highest common resolution ($8.5\arcsec$) with a circular beam (Fig. \ref{fig:400_700_8p5}).
For low-resolution images, we made \textit{uv-}tapered images using the \texttt{WSCLEAN} parameter \texttt{-taper-gaussian} corresponding to different beam sizes (e.g., $10\arcsec$, 15\arcsec,..., 30\arcsec). However, compared to the increasing RMS noise with increasing beam size, the additional diffuse emission recovery was minimal, and directly convolving the high-resolution image with a larger beam produced better results.
Here, we would like to note that the shortest baselines of band-3 and band-4 data are at $\sim50\lambda$ and $\sim70\lambda$, respectively. Since band-3 images with $>70\lambda$ \textit{uv-}data didn't result in any noticeable change, we proceeded with this minor difference.
For ease of reference, we have listed all the image properties in Table \ref{tab:image_prop}. 

\begin{deluxetable}{ccccc}
\tablecaption{Image properties \label{tab:image_prop}}
\tablecolumns{5}
\tablewidth{0pt}
\tablehead{
\colhead{Frequency} & \colhead{Name} & \colhead{Restoring} & \colhead{Briggs} &  \colhead{rms}\\ [-2ex]
\colhead{(MHz)} & \colhead{} & \colhead{beam} & \colhead{robust} &  \colhead{($\mu$Jy beam$^{-1}$)}}
\startdata
400 & IM1 & $7.5\arcsec \times 5.6\arcsec$ & 0   & $17$ \\
 & IM2 & $8.5\arcsec \times 8.5\arcsec$ & 0  & $20$ \\
 & IM3 & $15\arcsec \times 15\arcsec$ & 0 & $25$ \\
\hline
675 & IM4 & $8.4\arcsec \times 3.6\arcsec$ & 0 & $11$ \\
  & IM5 & $8.5\arcsec \times 8.5\arcsec$ & 0 & $15$ \\
  & IM6 & $15\arcsec \times 15\arcsec$ & 0 & $20$ \\
\enddata
\end{deluxetable}


\begin{figure}[!h]
    \centering
    \includegraphics[width=\columnwidth]{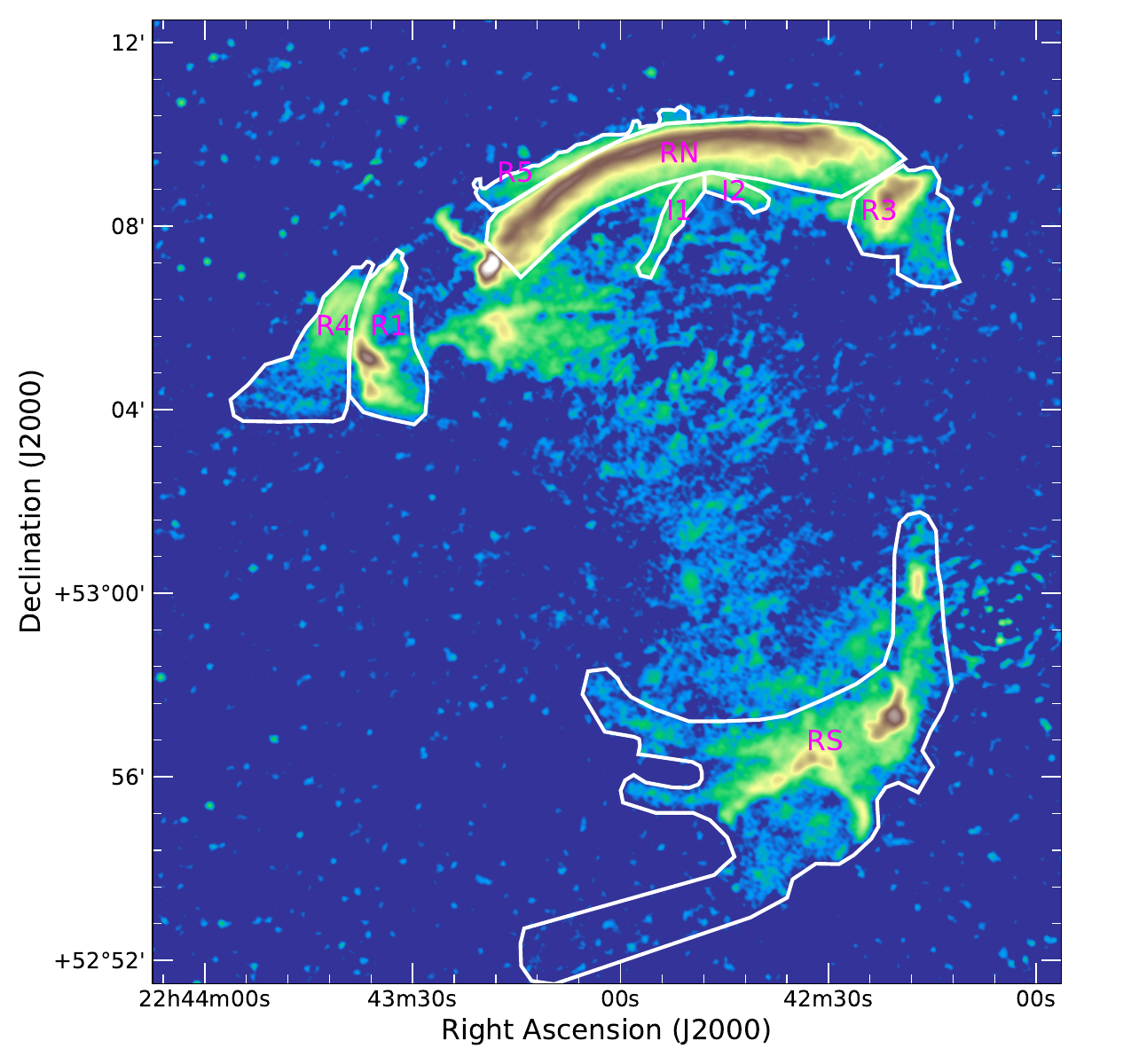}
    \caption{The uGMRT 400 MHz image, same as Fig. \ref{fig:400_700_8p5} (\textit{left panel}), with regions indicating individual diffuse radio sources over which integrated flux densities are estimated.}
    \label{fig:Regions}
\end{figure}

\begin{figure*}[!t]
    \centering
    \includegraphics[width=2.1\columnwidth]{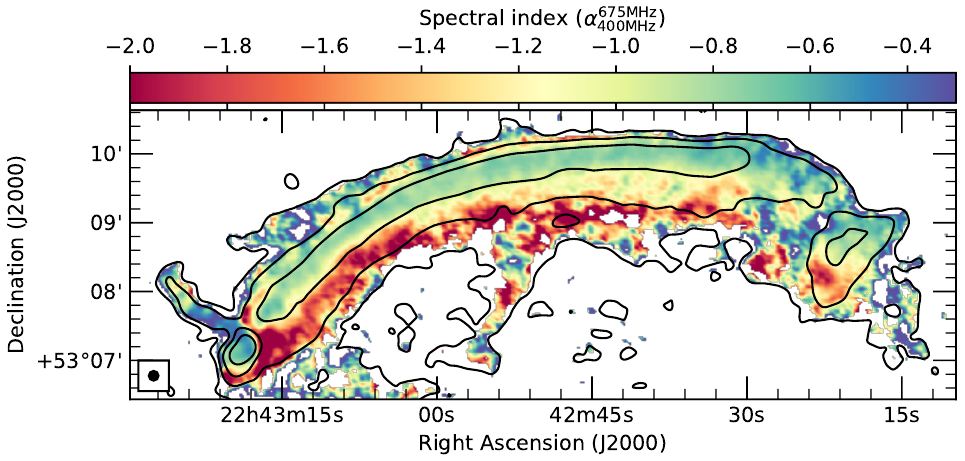}
    \caption{High-resolution ($8.5\arcsec$) spectral index map between 400 and 675 MHz of the RN, R5 and R3 relic. The images used to make this map are IM2 and IM5 in Table \ref{tab:image_prop}. The contours overlaid on the map correspond to the 400 MHz image (IM2), and are drawn at levels $[-1, 1, 4, 16,...] \times 4\sigma_{\rm rms}$.}
    \label{fig:RN_spix_map}
\end{figure*}

\begin{figure}[!t]
    \centering
    \includegraphics[width=\columnwidth]{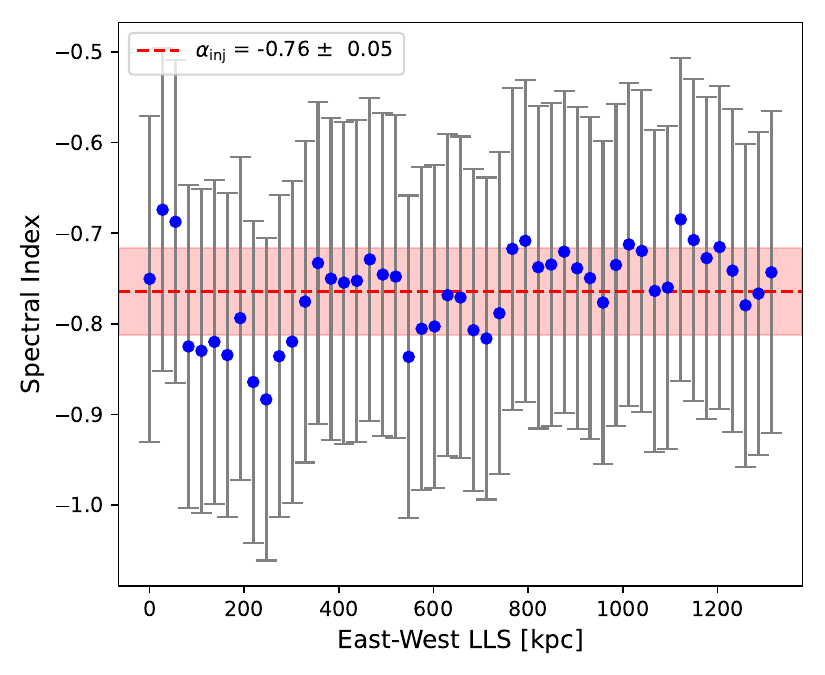}\\
    \vspace{-5pt}
    \includegraphics[width=\columnwidth]{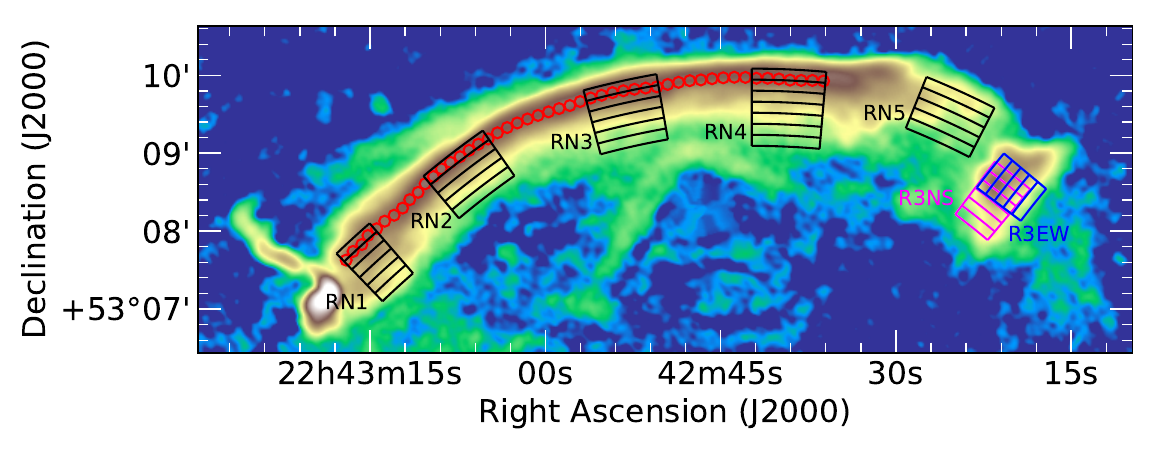}
    \caption{(\textit{Top}): East-west spectral index profile of the RN relic, extracted from the spectral index map in Fig. \ref{fig:RN_spix_map} within circular regions of the beam size, i.e., width = $8.5\arcsec$. (\textit{Bottom}): Zoomed-in 400 MHz image, shown in Fig. \ref{fig:400_700_8p5} (\textit{left panel}), focusing the northern relic. The red circular regions are used for the spectral index profile shown in the \textit{top panel}. The black sectors show the regions where we extracted spectral index profile across the RN relic width, whereas the blue and magenta sectors show the same for the R3 relic. These profiles are shown in Fig. \ref{fig:RN_pie_profile}. }
    \label{fig:RN_Regions_inj_spix}
\end{figure}

\begin{figure}[!t]
    \centering
    \includegraphics[width=\columnwidth]{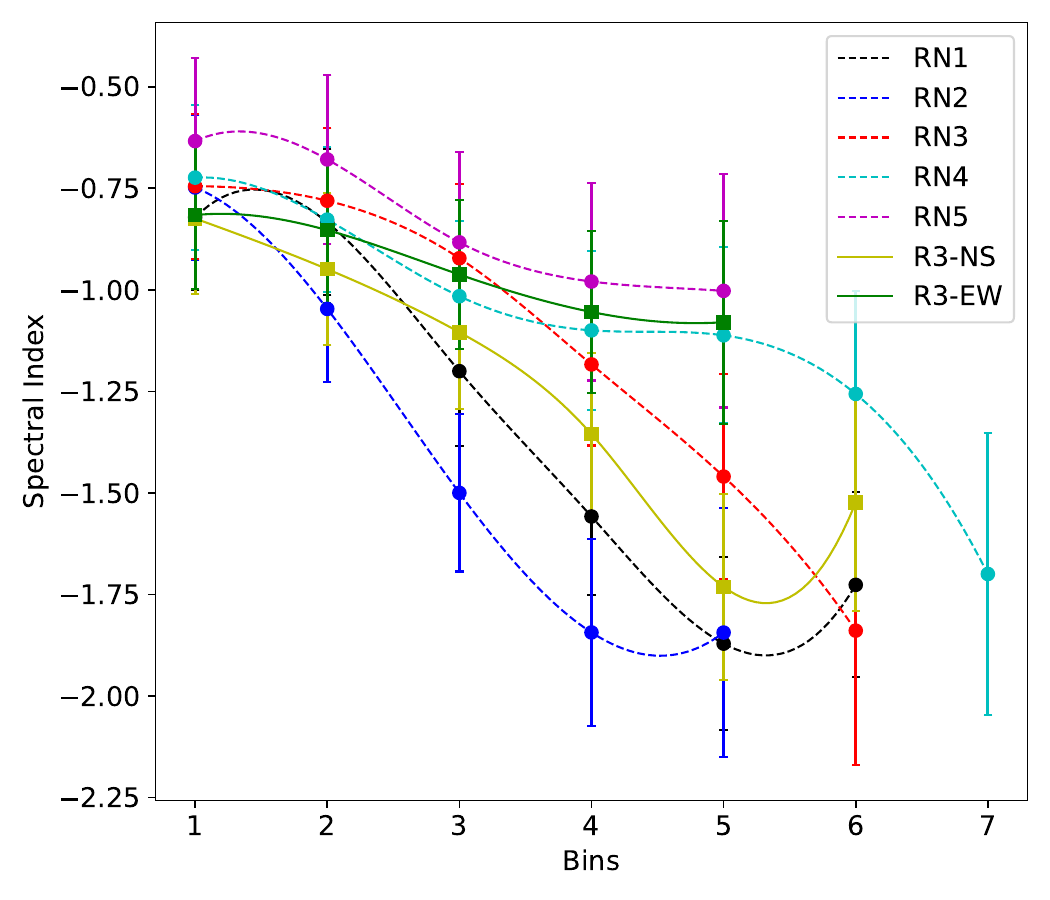}
    \caption{Spectral index profiles across the width of the RN and R3 relic. The regions within which these profiles are extracted are shown in Fig. \ref{fig:RN_Regions_inj_spix}, with the width of the rectangular boxes being $8.5\arcsec$.} 
    \label{fig:RN_pie_profile}
\end{figure}

\subsection{Flux density \& spectral index calculation} \label{subsec:flux_density_est}
In this work, flux density measurements are done on the RG subtracted images within the specified regions indicated accordingly (Fig. \ref{fig:Regions}). 
The flux density uncertainties are estimated using 
\vspace{-5pt}
\begin{equation}
    \sigma_{S} = \sqrt{(\sigma_{\mathrm{cal}} \times S)^2 + (\sigma_{\mathrm{rms}}\sqrt{N_{\mathrm{beam}}})^2}\,,
    \label{eq:flux_err}
\end{equation}

\noindent where $S$ is flux density, $\sigma_{\mathrm{cal}}$ and $\sigma_{\mathrm{rms}}$ are calibration uncertainty and image noise, respectively. The $N_{\mathrm{beam}}$ is the number of beams present within the said region.
In the flux density error estimation, we assumed $\sigma_{\mathrm{cal}}$ at 400 and 675 MHz to be $7\%$, and $6\%$ \citep[][]{Chandra2004ApJ...612..974C}, respectively.

As this study is based on two close-by frequencies with similar sensitivities, the spectral index values are very sensitive to small scale regional brightness fluctuations and missing flux, owing to the observational constraints. Consequently, a resolved spectral index map suffers from unrealistic values (i.e., too flat or too steep) in some regions, especially with low brightness. 
The spectral indices are derived using

\begin{equation}
    \alpha = \frac{\mathrm{ln} \Big(\frac{S_1}{S_2}\Big)}{\mathrm{ln} \Big(\frac{\nu_1}{\nu_2}\Big)}
    \label{eq:spec_index}
\end{equation}

\noindent where, $S_1$ and $S_2$ are flux densities corresponding to frequencies $\nu_1 = 400$ MHz and $\nu_2 = 675$ MHz, respectively. 
The spectral index uncertainties are calculated using 
\begin{equation}
    \Delta \alpha = \frac{1}{\ln\Big(\frac{\nu_1}{\nu_2}\Big)}\sqrt{\Bigg(\frac{\Delta S_1}{S_1}\Bigg)^2 + \Bigg(\frac{\Delta S_2}{S_2}\Bigg)^2}\,,
    \label{eq:spec_index_err}
\end{equation}
\noindent where $\Delta S_i = \sqrt{(\sigma_{\mathrm{cal}}^i \times S_i)^2 + (\sigma_{\mathrm{rms}}^i)^2}$ are total uncertainties in $S_i$. 

\begin{table}[]
\caption{Integrated flux densities and spectral indices of the diffuse sources}
\label{tab:int_spix}
\hskip-1.0cm\begin{tabular}{lccc}
\hline
\hline
Source & $S_{\rm 400~MHz}$ & $S_{\rm 675~MHz}$ & $\alpha_{\rm int}$ \\
	    & (mJy)	& (mJy)  &	  \\
\hline
RN     & $429.5 \pm 30.1$ & $262.6 \pm 15.8$ & $-0.94 \pm 0.18$ \\
RS     & $170.8 \pm 12.0$ & $96.7 \pm 5.8$ & $-1.09 \pm 0.18$ \\
R1     & $53.2 \pm 3.7$ & $36.0 \pm 2.2$ & $-0.75 \pm 0.18$ \\
R3     & $52.1 \pm 3.7$ & $33.3 \pm 2.0$ & $-0.86 \pm 0.18$ \\
R4     & $21.9 \pm 1.6$ & $16.1 \pm 1.0$ & $-0.6 \pm 0.2$ \\
R5     & $9.9 \pm 0.7$ & $7.3 \pm 0.5$ & $-0.6 \pm 0.2$ \\
I1     & $8.1 \pm 0.6$ & $3.9 \pm 0.2$ & $-1.39 \pm 0.18$ \\
I2     & $5.4 \pm 0.4$ & $2.5 \pm 0.2$ & $-1.44 \pm 0.18$ \\
\hline
\end{tabular}
\end{table}

\section{Northern Relics} \label{sec:NR}
The northern relic region is composed of a number of diffuse radio structures spreading over $\sim 3.5$ Mpc (largest linear size or LLS). The most prominent source among them is obviously the `North Relic' RN or the `Sausage' relic. Then the sources directly connected with this are relic R5, R3 and source I1 \& I2 in the north, west and south of the RN, respectively. On the eastern side, relic R1 and R4 shows different morphological features. Detailed discussion on these sources are presented in the following sections.

\subsection{Sausage or North Relic: RN} \label{subsec:RN}
The north relic RN, or the `Sausage' relic is detected at both 400 and 675 MHz uGMRT observations in detail (Fig. \ref{fig:400_700_8p5}). The well-known arc shape of the relic is clearly visible, with the northern side being the brightest with a sharp edge towards the north and a gradual decline in surface brightness towards the south. 
On the eastern side, the relic touches the source H (an AGN), then R5 in the north, followed by I1 \& I2 in the south and finally connects with the relic R3 with a faint bridge in the west (Fig. \ref{fig:Labels}).

The projected linear size of the relic recovered at both frequencies is similar, spanning $\sim 1.8$ Mpc in the east-west direction with a narrow width of $\sim 180$ kpc. This linear size does not include the relic R3 (see Fig. \ref{fig:Regions}). The integrated flux densities of the relic are $S_{400} = 429.5 \pm 30.1$ and $S_{675} = 262.6 \pm 15.8$ mJy (Table \ref{tab:int_spix}). The flux density values are estimated within the RN region indicated in Fig. \ref{fig:Regions}. The integrated spectral index of the RN relic comes out to be $\alpha_{\rm int} = -0.94 \pm 0.18$, which is consistent to within the uncertainties with previous estimations ($\alpha_\mathrm{0.61GHz}^\mathrm{2.3GHz} = -1.08 \pm 0.05$, \citealt{vanWeeren2010Sci...330..347V}; $\alpha_\mathrm{0.153GHz}^\mathrm{2.3GHz} = -1.06 \pm 0.04$, \citealt{Stroe2013A&A...555A.110S}; $\alpha_\mathrm{0.145GHz}^\mathrm{2.3GHz} = -1.11 \pm 0.04$, \citealt{Hoang2017MNRAS.471.1107H}; $\alpha_\mathrm{0.153GHz}^\mathrm{8.35GHz} = -0.90 \pm 0.04$ \citealt{Kierdorf2017A&A...600A..18K}; $\alpha_\mathrm{1.5GHz}^\mathrm{3.0GHz} = -1.19 \pm 0.05$, \citealt{DiGennaro2018ApJ...865...24D}; $\alpha_\mathrm{0.145GHz}^\mathrm{18.6GHz} = -1.12 \pm 0.03$ \citealt{Loi2020MNRAS.498.1628L}). In the spectral index map (Fig. \ref{fig:RN_spix_map}), we see the spectral index near the northern edge at around $-0.8$, gradually steepening downstream, ranging from $-1.2$ to $\lesssim -1.5$ depending upon the region. For quantitative analysis, we have produced different spectral profiles across the radio relic. Following \citet{DiGennaro2018ApJ...865...24D} as reference, we chose to extract spectral index profiles corresponding to regions RN1 to RN5 (see Fig. \ref{fig:RN_Regions_inj_spix} {\it bottom}). From Fig. \ref{fig:RN_pie_profile}, we observe that spectral index gradient across the relic width is highest in regions RN1 \& RN2, followed by RN3, whereas RN4 \& RN5 shows much more gradual decline.
This indicates radiative cooling of the shock downstream region associated with inverse Compton (IC) and synchrotron loss. However, the reason for different cooling rates in different downstream regions is not clear. \citet{DiGennaro2021ApJ...911....3D} reported polarized emission and magnetic field variation along the relic from the east to west. Therefore, a possible explanation can be that a higher magnetic field will result in greater cooling in the east of the relic and vice versa.

To calculate Mach number of the `Sausage' relic corresponding to these new observations, we have first estimated the injection spectral index. To do that, we have plotted a spectral index profile along the northern edge of the relic across the relic length. To avoid mixing of different electron populations as much as possible, we chose the sample region to be the same as the beam size (i.e., $8.5\arcsec$). The Fig. \ref{fig:RN_Regions_inj_spix} ({\it top}) shows that the spectral indices remains roughly similar throughout most of the relic length. Using equation
\begin{equation}
    \mathcal{M} = \sqrt{\frac{2\alpha_{\rm inj} - 3}{2\alpha_{\rm inj} + 1}}
    \label{eq:Mach_number}
\end{equation}

\noindent \citep[e.g.,][]{Drury1983RPPh...46..973D,Blandford1987PhR...154....1B}{}{} the Mach number corresponding to the $\alpha_{\rm inj} = -0.76 \pm 0.05$ comes out to be $\mathcal{M} = 2.9_{-0.2}^{+0.3}$, which is in agreement with the previous studies (see Table \ref{tab:Mach_numbers}). Here, the mean scatter in $\alpha_{\rm inj}$ is derived following \citet{Cassano2013ApJ...777..141C,vanWeeren2016ApJ...818..204V}. The observed slight variation in the injection spectral indices across the length of the RN relic is possibly related to the variation in the Mach numbers, magnetic fields, electron distribution, etc. and the compelling evidence for that may be the discovery of filaments/sheets in the RN relic by \citet{DiGennaro2018ApJ...865...24D}.

\subsection{Relic R5} \label{subsec:R5}
\begin{figure}
    \centering
    \includegraphics[width=\columnwidth]{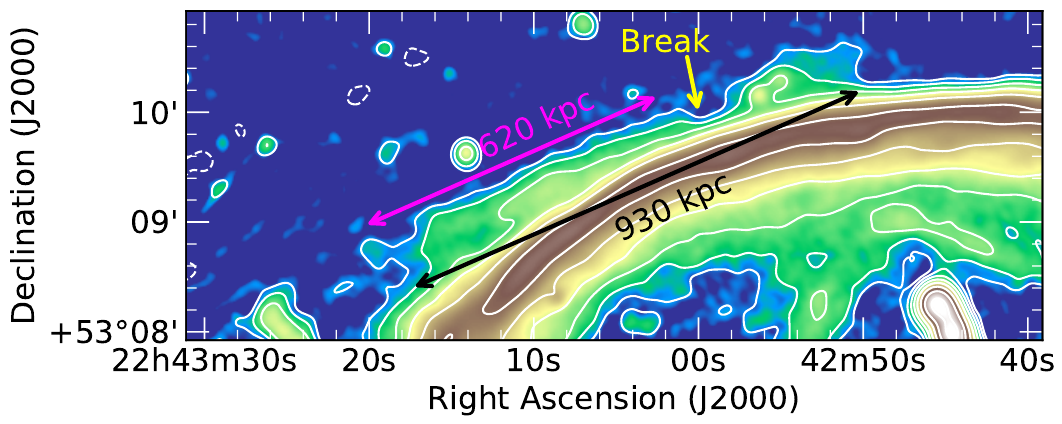}
    \caption{Zoomed-in 400 MHz image in Fig. \ref{fig:400_700_8p5} (compact sources un-subtracted), focusing the relic R5.}
    \label{fig:R5_length}
\end{figure}

\begin{table*}[]
\caption{Spectral Indices and Mach Numbers for the Relics}
\label{tab:Mach_numbers}
\begin{center}
\resizebox{2\columnwidth}{!}{%
\hskip-0.8in\begin{tabular}{lcccccc}
\hline
\hline
Source & $\alpha_{\rm inj}$ & $\mathcal{M}$ & $\mathcal{M}_{\rm Stroe}$ & $\mathcal{M}_{\rm Hoang}^{c}$ & $\mathcal{M}_{\rm DiGennaro}^{d}$ & $\mathcal{M}_{\rm X-ray}$ \\
\hline
RN & $-0.76 \pm 0.05$ & $2.9_{-0.2}^{+0.3}$ & $2.9_{-0.13}^{+0.10\ b}$ & $2.7_{-0.3}^{+0.6}$ & $2.58 \pm 0.17$ & $2.7_{-0.4}^{+0.7}$ \\
R1 & $-0.71 \pm 0.18$ & $3.2_{-0.8}^{+5.0}$ &  & $2.4_{-0.2}^{+0.4\ \dagger}$ & $2.69 \pm 0.06$ & $2.6_{-0.2}^{+0.6\ c}$ \\
R3 & $-0.78 \pm 0.18$ & $2.8_{-0.5}^{+1.7}$ &  &  &  &  \\
R4 & $-0.82 \pm 0.20$ & $2.7_{-0.5}^{+1.5}$ &  & $2.5_{-0.2}^{+0.2\ \ddagger}$ & $2.57 \pm 0.12$ &  \\
R5 & $-0.80 \pm 0.36$ & $2.8_{-0.8}^{*}$ &  &  & $2.13 \pm 0.55$ &  \\
RS1 & $-0.89 \pm 0.25$ & $2.5_{-0.4}^{+1.4}$ &  &  &  &  \\
RS3 & $-0.97 \pm 0.32$ & $2.3_{-0.4}^{+1.5}$ &  &  &  &  \\
RS4 & $-1.17 \pm 0.56$ & $2.0_{-0.4}^{+2.3}$ &  &  &  &  \\
RS & $-1.02 \pm 0.40$ & $2.2_{-0.4}^{+2.1}$ & $2.8_{-0.2}^{+0.2\ a}$ & $1.9_{-0.2}^{+0.3}$ & $2.10 \pm 0.08$ & $1.7_{-0.3}^{+0.4\ e}$ \\
\hline
\end{tabular}%
}
\end{center}
{{\bf Notes.} $^{a}$\citet{Stroe2013A&A...555A.110S}, $^{b}$\citet{Stroe2014MNRAS.445.1213S}}, $^{c}$\citet{Hoang2017MNRAS.471.1107H}, $^{d}$\citet{DiGennaro2018ApJ...865...24D}, $^{e}$\citet{Akamatsu2015A&A...582A..87A}
\newline
$^\dagger$Derived from R1W $\alpha_{\rm inj} = -0.92 \pm 0.12$ \citep{Hoang2017MNRAS.471.1107H} \\
$^\ddagger$Derived from R1E $\alpha_{\rm inj} = -0.89 \pm 0.08$ \citep{Hoang2017MNRAS.471.1107H}\\
$^{*}$Upper limit blows up to infinity.
\end{table*}

The presence of additional faint diffuse radio emission at the north-east of the Sausage relic was first reported by \citet{DiGennaro2018ApJ...865...24D}, labeled as relic R5. In this work, we re-confirm the presence of this source as it is observed clearly in both 400 and 675 MHz images (Fig. \ref{fig:400_700_8p5}), showing striking resemblance with Fig. 3, \citet{DiGennaro2018ApJ...865...24D}. 
In fact, in hindsight, we see traces of this emission in the WSRT image by \citet{vanWeeren2010Sci...330..347V} as well.
The relic R5 is fairly straight without any visible bend, unlike the other relics in this cluster. 

The relic starts with $\sim 100$ kpc width at the eastern end and gradually narrows westward, eventually merging with relic RN. The existence of a break towards the western end of the R5 relic also matches with \citet{DiGennaro2018ApJ...865...24D} VLA observations. The linear size of the R5 from the east up to the break is $\sim 620$ kpc (Fig. \ref{fig:R5_length}), similar to the measurement reported by \citet{DiGennaro2018ApJ...865...24D}. However, if we include the diffuse emission beyond the break point as part of the R5 relic, then the total length of the relic becomes $\sim 930$ kpc. 

The integrated flux density values measured within the R5 region indicated in Fig. \ref{fig:Regions} are $S_{400} = 9.9 \pm 0.7$ and $S_{675} = 7.3 \pm 0.5$ mJy, resulting in the spectral index of $\alpha_{\rm int} = -0.6 \pm 0.2$ (Table \ref{tab:int_spix}). This is much flatter than the \citet{DiGennaro2018ApJ...865...24D} reported value ($\alpha_\mathrm{1.5GHz}^\mathrm{3.0GHz} = -0.93 \pm 0.12$). 
The most possible cause for this flatness can be the missing flux at 400 MHz observations, as we notice the R5 relic to be slightly better recovered at 675 MHz image. Similar bias is also observed in other relics with faint diffuse component, discussed later. 
We measured injection spectral index of the R5 within a rectangular box of width $8.5\arcsec$ along the northern edge of the relic to be $\alpha_{\rm inj} = -0.8 \pm 0.36$. 
This results into a Mach number of $\mathcal{M} = 2.8_{-0.8}$, which has large uncertainties and therefore not that useful but reported here for completeness.

\begin{figure*}[!ht]
    \centering
    \begin{tabular}{cc}
    \includegraphics[width=\columnwidth]{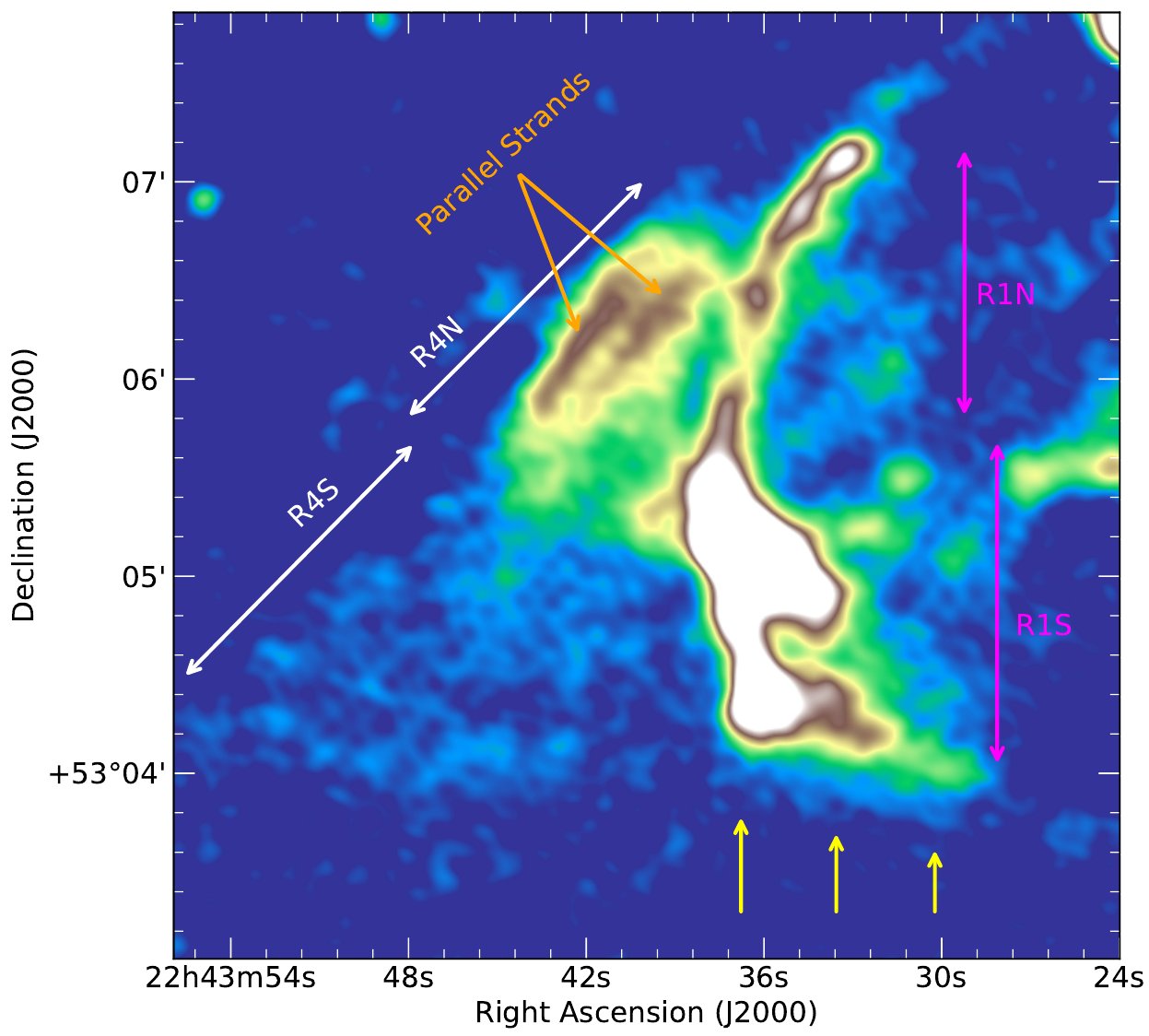} & 
    \includegraphics[width=\columnwidth]{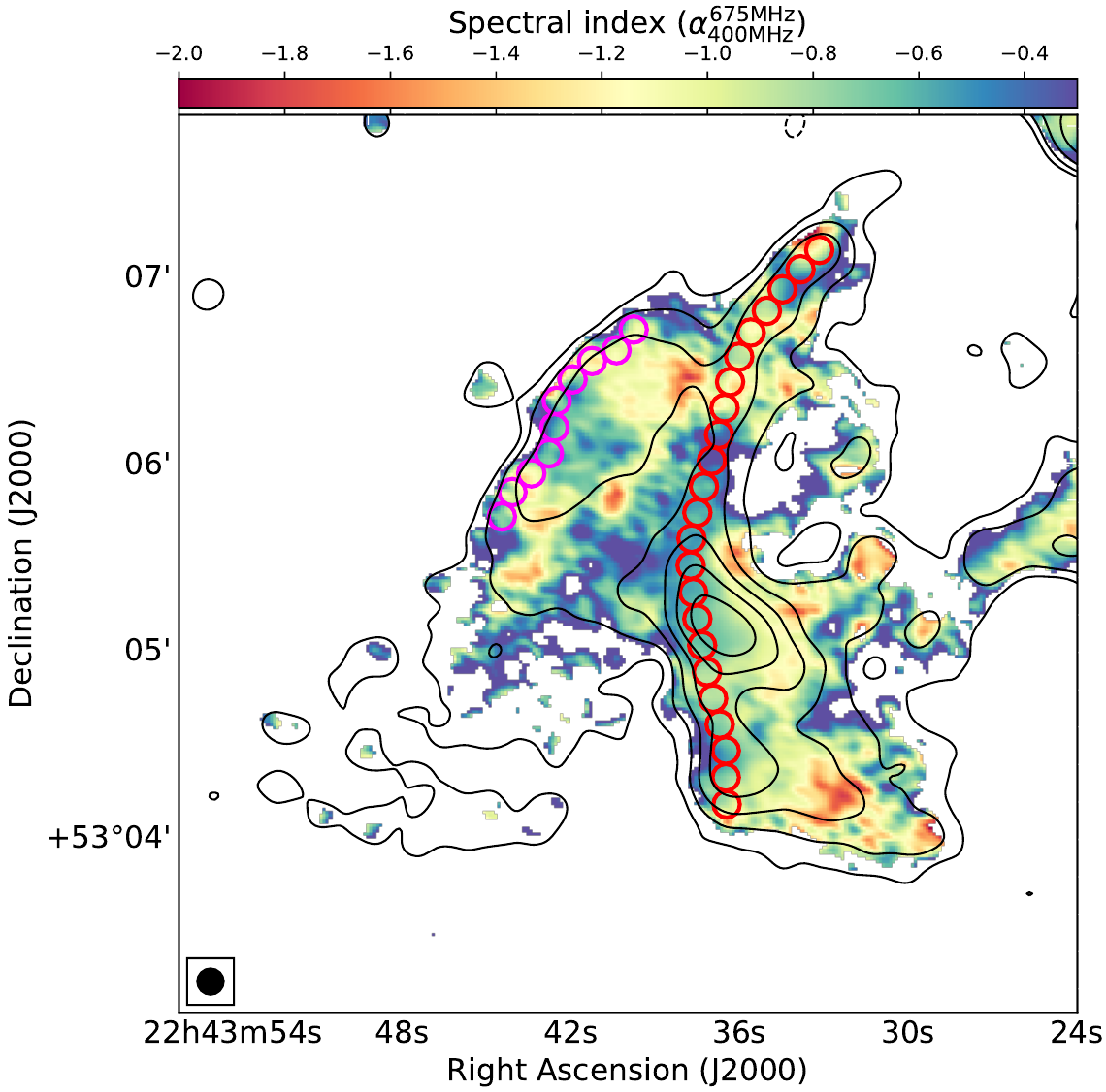}
    \end{tabular}
    \caption{(\textit{Left}): Zoomed-in 400 MHz image (Fig. \ref{fig:400_700_8p5} \textit{left panel}) in linear scale, showing the different features of the R1 and R4 relic. (\textit{Right}): High-resolution ($8.5\arcsec$) spectral index map between 400 and 675 MHz of the R1 and R4 relic. The images used to make this map are IM2 and IM5 in Table \ref{tab:image_prop}. The contours overlaid on the map correspond to the 400 MHz image (IM2), and are drawn at levels $[-1, 1, 2, 4, 8,...] \times 3\sigma_{\rm rms}$. The circular regions of beam width are used to extract spectral index and brightness profiles of the R1 (red) and R4 (magenta) relic. These profiles are presented in Fig. \ref{fig:R1_R4_inj_spix}.}
    \label{fig:R1_R4_linear_spix}
\end{figure*}

\begin{figure*}[!ht]
    \centering
    \begin{tabular}{cc}
    \includegraphics[width=\columnwidth]{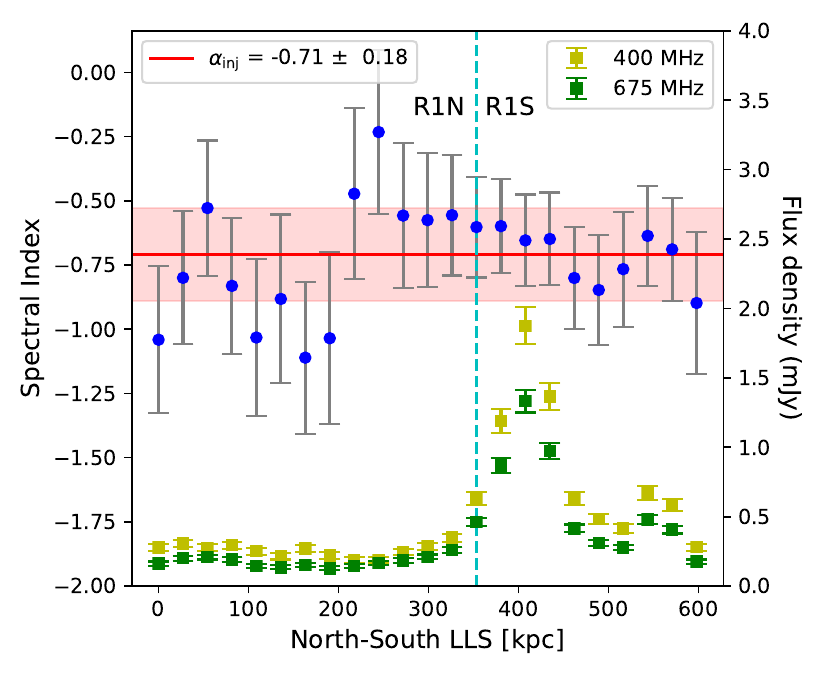} & 
    \includegraphics[width=\columnwidth]{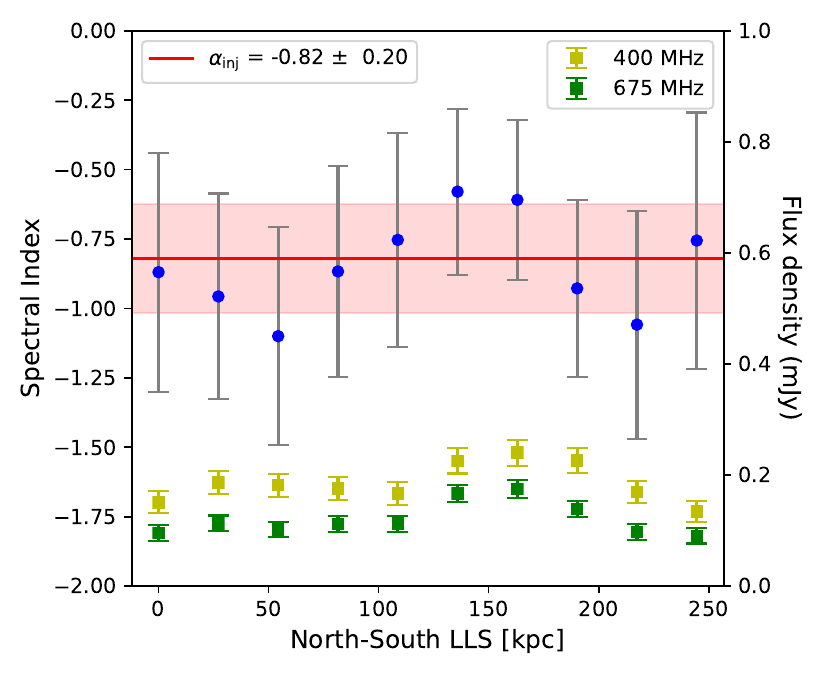}
    \end{tabular}
    \caption{North-south spectral index and brightness profiles of the R1 (\textit{left}) and R4 (\textit{right}) relic, corresponding to the regions shown in Fig. \ref{fig:R1_R4_linear_spix}.}
    \label{fig:R1_R4_inj_spix}
\end{figure*}

\subsection{East Relic: R1} \label{subsec:R1}
On the east of the RN relic, a patch of diffuse emission with complex morphology is detected at both 400 and 675 MHz, labelled as R1 and R4 (Fig. \ref{fig:400_700_8p5}). Positioned at similar projected distance as RN from the cluster center, \citet{Stroe2013A&A...555A.110S} suggested R1 to be related to a common merger event.  
The relic R1 consists of a bright thin filament (width $\sim 39$ kpc) like structure stretching in the north-south direction for $\sim 630$ kpc, with a sharp edge towards the east, and a faint non-filamentary diffuse component extending towards the west. 
The similar arc-like morphology (as RN) of R1 relic suggest it being seen edge-on \citep{Stroe2013A&A...555A.110S}.
The full size of the relic R1 is $\sim 330 \times 740$ kpc. The integrated flux densities of the R1 relic are $S_{400} = 53.2 \pm 3.7$ and $S_{675} = 36.0 \pm 2.2$ mJy, resulting in the spectral index of $\alpha_{\rm int} = -0.75 \pm 0.18$ (Table \ref{tab:int_spix}). We measured injection spectral index along the bright arc filament following the similar method as described in Sect. \ref{subsec:RN} to be $\alpha_{\rm inj} = -0.71 \pm 0.18$ (Fig. \ref{fig:R1_R4_inj_spix} {\it left panel}). 
Therefore, the corresponding Mach number comes out to be $\mathcal{M} = 3.2_{-0.8}^{+5.5}$. 
This is much higher than the previously reported values, and the associated large uncertainties makes it less useful but reported here for completeness.

Morphologically, the R1 relic can be roughly divided into northern (R1N) and southern (R1S) half at the junction of the bright spot in the middle (Fig. \ref{fig:R1_R4_linear_spix} {\it left panel}). 
The R1N region consists of a thin arc-like bright filament relic followed by a faint diffuse emission resembling a typical relic downstream region. However, because of low signal-to-noise (S/N) recovery of this downstream diffuse emission, we are unable to probe the existence of spectral gradient (if any) in this region.
On the other hand, the R1S region has a more complex morphology, with bright threads forking from the main north-south arc filament towards the cluster center, accompanied by similar downstream faint diffuse emission. The R1S region is also considerably brighter compared to R1N. 
The spectral index map in Fig. \ref{fig:R1_R4_linear_spix} ({\it right panel}) shows the presence of a spectral gradient from the position of the relic filament (possible position of shock front) towards the west in the downstream region, as was previously reported by \citet{Hoang2017MNRAS.471.1107H,DiGennaro2018ApJ...865...24D}.
Furthermore, we also observe a rapid brightness drop in R1S in the southern direction at both frequencies, forming an edge-like feature (Fig. \ref{fig:R1_R4_linear_spix} {\it left panel}) in contrast with a more gradual decease in the western direction. 
In hindsight, this feature was also observed in the images by \citet{Hoang2017MNRAS.471.1107H} and \citet{DiGennaro2018ApJ...865...24D}. From the observed morphology of this feature, it seems like the old electron population bubble encountered some kind of pressure barrier restricting diffusion in that direction and the subsequent merger shock (associated with the relic R1) pressure resulted in the observed thread-like structure along that barrier, as well as the sharp brightness drop. 

Besides the division of the relic based on the apparent brightness distribution (i.e., R1N and R1S), we also observe weak sub-structure in the injection spectral index profile. In Fig. \ref{fig:R1_R4_inj_spix} ({\it left panel}) we see that the northern $\sim 200$ kpc of the relic seems slightly steeper compared to the rest of the southern part. The most likely cause of this can be the presence of multiple finer filaments \citep[similar to the RN relic in Fig. 7 \& 8,][]{DiGennaro2018ApJ...865...24D}{}{} with varied Mach number, magnetic fields, electron distribution, etc. as was previously observed in RN relic. However, the radio observational data used in this work lacks sufficient resolution to explore this further.

\subsection{East Relic: R4} \label{subsec:R4}
The diffuse radio source on the east of the relic R1, labelled as R4 (Fig. \ref{fig:Labels}), starts from the eastern edge of the R1 relic and extends further east in a similar manner as was previously observed in relic R5 (see Fig. \ref{fig:R5_length}). However, unlike the R5, which stretched parallel to the RN relic tangent, R4 extends obliquely at $\sim 45^\circ$ angle to the relic R1. Therefore, the relic width increases from the narrow $\sim 80$ kpc at the north to the astonishing $\sim 490$ kpc at the southern end, with the largest linear size (LLS) being $\sim 840$ kpc. 

There is a clear brightness gradient in the north-south direction which can roughly be divided into north R4N and south R4S region (see Fig. \ref{fig:R1_R4_linear_spix} {\it left panel}). A possible cause can be the difference in electron density between these regions.
Moreover, we also observe two `parallel strands' of width $\sim 48$ kpc closely aligned with the north-eastern edge of the R4 relic in the R4N region at both 400 and 675 MHz images, indicating their possible connection with the shock front responsible for the R4 relic itself. 
On the other hand, the R4S region has smooth brightness distribution without any filamentary features. 
The double-strand feature in the R4N region can be compared to the similar features discovered in the `Toothbrush' relic by \citet{Rajpurohit2018ApJ...852...65R}.  
Due to the sensitivity and resolution limitation, any further investigation on these with the current data is unreliable.

The integrated flux densities of the whole R4 relic estimated within the region indicated in Fig. \ref{fig:Regions} are $S_{400} = 21.9 \pm 1.6$ and $S_{675} = 16.1 \pm 1.0$ mJy. Therefore, the integrated spectral index comes out to be $\alpha_{\rm int} = -0.6 \pm 0.2$ (Table \ref{tab:int_spix}). As we see that the low surface brightness component of the relic is slightly better recovered at 675 MHz image (see Fig. \ref{fig:400_700_8p5}), this might have resulted in the flat spectral index. 
To explore this further, we derived integrated spectral index for R4N and R4S separately, which came out to be $-0.72 \pm 0.18$ and $-0.49 \pm 0.18$, respectively. This result supports the earlier assumption on the flat spectral index bias. 

To derive the Mach number associated with the relic R4, we estimated injection spectral index along the north-eastern edge of the relic (see Fig. \ref{fig:R1_R4_inj_spix} {\it right panel}). We get the Mach number corresponding to $\alpha_{\rm inj} = -0.82 \pm 0.20$ (Fig. \ref{fig:R1_R4_inj_spix} {\it right panel}) to be $\mathcal{M} = 2.7_{-0.5}^{+1.5}$, which is consistent to within the uncertainties with the previous estimations by \citet{Hoang2017MNRAS.471.1107H,DiGennaro2018ApJ...865...24D}. 
\citet{Hoang2017MNRAS.471.1107H} suggested that the R1 and R4 relics trace a complex shock wave that is moving away from the cluster center (re-)accelerating two electron clouds along its way. 
On the other hand, morphological and $\alpha_{\rm inj}$ similarity between the R4 and R5 relics as well as symmetry with the southern RS5 relic may suggest a different scenario where these two relics are associated with the same merging shock paired with the RS5 relic (see Fig. \ref{fig:shock_arcs}). However, the current data is insufficient to allow exploring these questions.

\begin{figure*}
    \centering
    \begin{tabular}{cc}
    \includegraphics[width=\columnwidth]{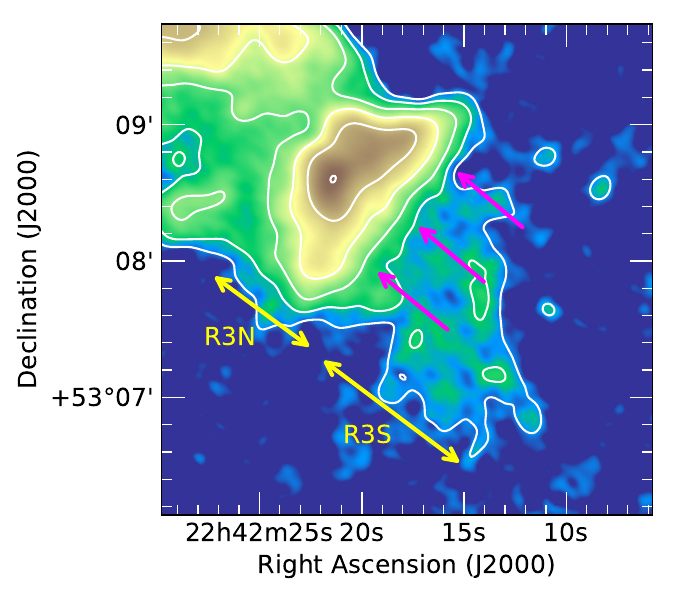}  & 
    \includegraphics[width=\columnwidth]{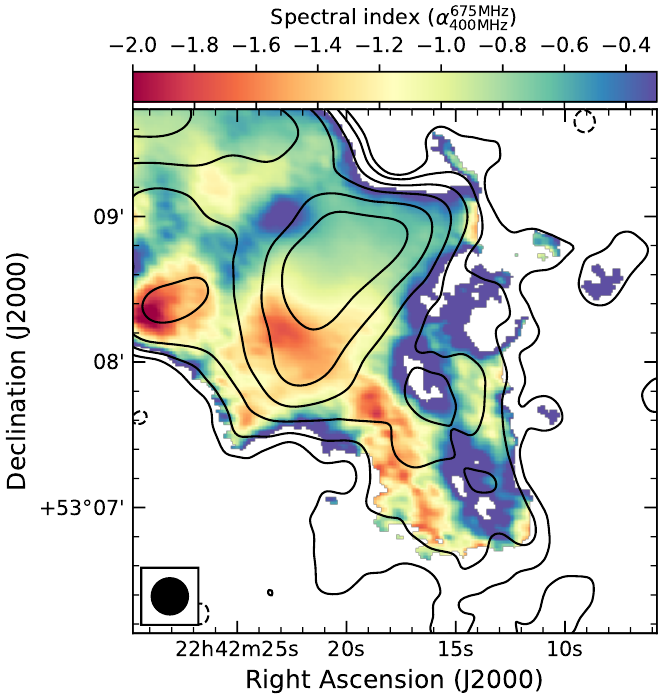}\\
    \end{tabular}
    \caption{(\textit{Left}): Zoomed-in 400 MHz image, shown in Fig. \ref{fig:400_700_8p5} (\textit{left panel}), focusing the relic R3. (\textit{Right}): Low-resolution ($15\arcsec$) spectral index map of the R3 relic between 400 and 675 MHz. The images used to make this map are IM3 and IM6 in Table \ref{tab:image_prop}. The contours overlaid on the map correspond to the 400 MHz image (IM3), and are drawn at levels $[-1, 1, 2, 4, 8,...] \times 3\sigma_{\rm rms}$.}
    \label{fig:R3_maps}
\end{figure*}

\begin{figure}
    \centering
    \includegraphics[width=\columnwidth]{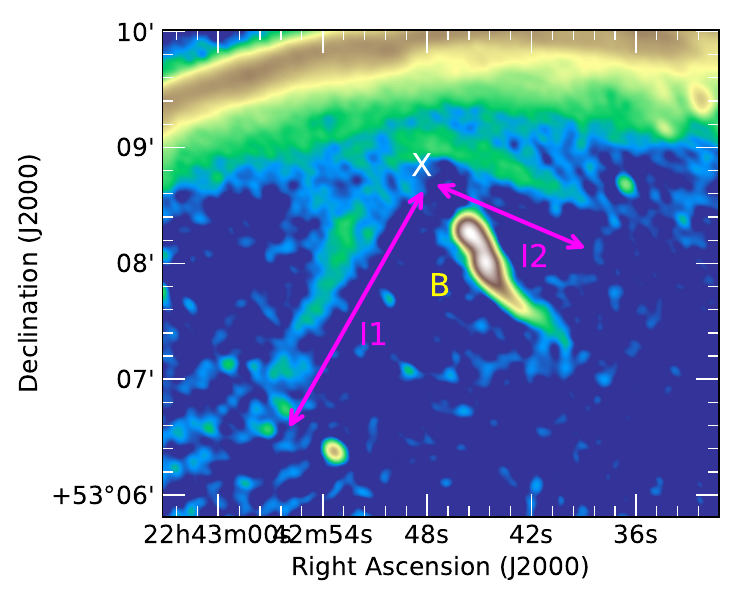}
    \caption{Zoomed-in 400 MHz image in Fig. \ref{fig:400_700_8p5} (compact sources un-subtracted), focusing the source I. The white `X' marks the location of a cluster member galaxy [DJS2015] J224248.254+530849.495 \citep{Dawson2015ApJ...805..143D}.}
    \label{fig:I1_I2}
\end{figure}

\subsection{West Relic: R3} \label{subsec:R3}
On the western end of the RN relic, a patch of diffuse radio emission of complex morphology is labelled as relic R3 (Fig. \ref{fig:Labels}).
Before \citet{Hoang2017MNRAS.471.1107H} who first labelled this source as R3, it was considered a part of the RN relic, and was not studied separately. 
The R3 relic is connected with the RN relic with a low surface brightness bridge (Fig. \ref{fig:400_700_8p5}).
In our 400 and 675 MHz images, this source is detected at a larger scale with an additional low surface brightness component, which was only detected in previous reports at a very low-resolution \citep[$\geq 25\arcsec,$][]{Hoang2017MNRAS.471.1107H,DiGennaro2018ApJ...865...24D}{}{}. We measure the size of the relic R3 to be $\sim 400 \times 540$ kpc, which can roughly be divided into a northern bright component R3N and southern faint component R3S. 
The transition from the bright R3N to the faint R3S has a sharp brightness decrease, as is evident from the tightness of the contours at both frequencies (Fig. \ref{fig:R3_maps}). 

The overall integrated flux densities of the R3 relic are $S_{\rm 400} = 52.1 \pm 3.7$ and $S_{\rm 675} = 33.3 \pm 2.0$ mJy, resulting in a spectral index of $\alpha_{\rm int} = -0.86 \pm 0.18$ (Table \ref{tab:int_spix}). Since R3 has two distinct components, we also measured integrated flux densities separately. Therefore, the flux densities of R3N are estimated to be $S_{\rm 400} = 44.4 \pm 3.1$ \& $S_{\rm 675} = 27.6 \pm 1.7$ mJy with $\alpha = -0.91 \pm 0.18$, and of the R3S to be $S_{\rm 400} = 7.1 \pm 0.5$ \& $S_{\rm 675} = 5.1 \pm 0.3$ mJy with $\alpha = -0.63 \pm 0.18$. 
We observe a similar flattening trend for the low surface brightness component of the R3 relic, suffering from total flux recovery, as was previously seen in the R4 relic in Sect. \ref{subsec:R4}. 

Fig. \ref{fig:RN_spix_map} shows a resolved spectral index map of the relic R3, but only of the bright R3N region, as the R3S region has poor S/N at this resolution ($8.5\arcsec$). We see that R3 also has similar spectral index gradient across its width, as was previously observed in relic RN, in both R3N and R3S regions (Fig. \ref{fig:R3_maps} \textit{right}). 
A spectral index profile across its width is plotted in Fig. \ref{fig:RN_pie_profile}, which shows a strong spectral gradient from $-0.8$ to $-1.7$ towards the cluster center. 
On the other hand, a weaker spectral gradient is seen in the east-west direction from $-0.8$ to $-1.05$. These results are in agreement with the reports by \citet{DiGennaro2018ApJ...865...24D}.  
Measuring the injection spectral index corresponding to a box region of width $8.5\arcsec$ in the R3N region, we get $\alpha_{\rm inj} = -0.78 \pm 0.18$. Therefore, the derived Mach number is $\mathcal{M} = 2.8_{-0.5}^{+1.7}$. 
Hence, in light of the spectral index profiles across the relic width (Fig. \ref{fig:RN_pie_profile}) and the derived Mach numbers, it can be speculated that the shock responsible for both RN and R3 relics is the same. The observed differences between them may be associated with the differences in seed electron densities and/or magnetic fields in these regions \citep[e.g., the differnce in polarization between eastern and western sides of the RN relic,][]{DiGennaro2021ApJ...911....3D}{}{}. 
However, we also note that the RN relic shows similar characteristics from RN1 to RN4, but falters at around region RN5 in both the brightness (Fig. \ref{fig:Labels}) and spectral index map (Fig. \ref{fig:RN_spix_map}). 
The high-resolution images in \citet{DiGennaro2018ApJ...865...24D} show filament/sheet-like features ending just before the RN5 region. 
Moreover, RN5 seems more similar to the R3 relic, with its patchy spectral index map (Fig. \ref{fig:RN_spix_map}). Furthermore, the combined RN5 and R3 relics can also be related with the relic RS3 symmetrically, as a possible merger event shock pair (see Fig. \ref{fig:shock_arcs}). In this scenario, the RN relic from RN1 to RN4 is associated with one shock, and the RN5 and R3 are associated with another concurrent merger shock wave. Further detailed investigation with better data is needed to provide any conclusive statements in this regard. 

\begin{figure*}
    \centering
    \includegraphics[width=2.1\columnwidth]{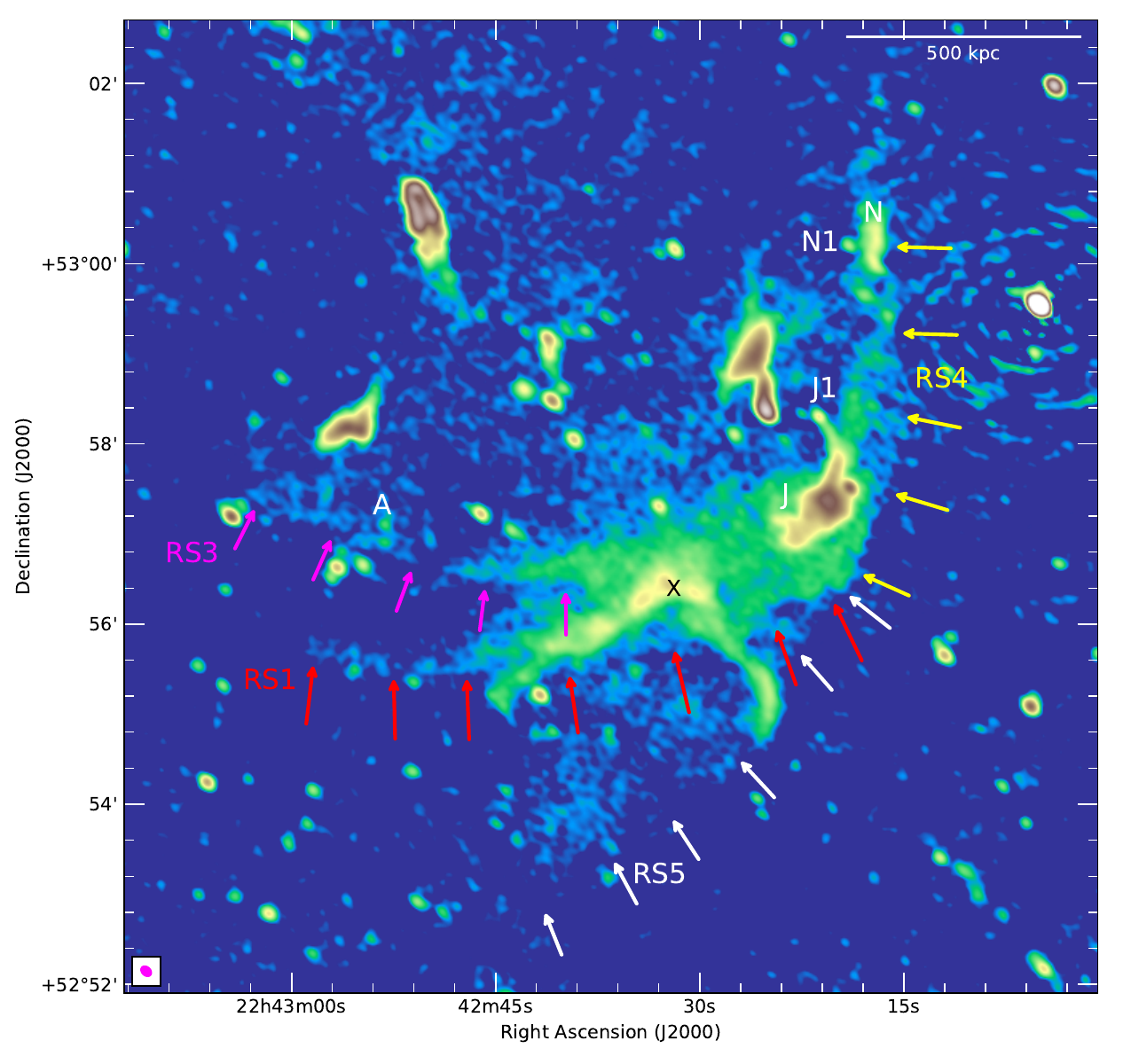}
    \caption{Zoomed-in high-resolution 400 MHz image (IM1), shown in Fig. \ref{fig:Labels}, focusing the southern relic. The arrows tracing the component relic extents. The black `X' marks the location of the optical counterpart of a cluster member galaxy [DJS2015] J224231.818+525622.413 \citep{Dawson2015ApJ...805..143D}.}
    \label{fig:RS_arrows}
\end{figure*}

\subsection{Source I1 \& I2}
Immediately south of the RN relic and north of the tailed radio galaxy B, the diffuse sources I1 and I2 are detected at both 400 and 675 MHz, connected with the downstream region of the RN relic (Fig. \ref{fig:400_700_8p5}). The source I1 was previously labelled as source I by \citet{Stroe2013A&A...555A.110S,Hoang2017MNRAS.471.1107H,DiGennaro2018ApJ...865...24D}. Here, we report the presence of a similar source I2 extended in the perpendicular direction, originating from the same RN downstream region as I1 (see Fig. \ref{fig:Labels}, \ref{fig:I1_I2}). Although not fully defined, a trace of the source I2 was already detected in the LOFAR image by \citet{Hoang2017MNRAS.471.1107H}, which they considered together with the I1 as an arc-like object with a shock morphology. Surprisingly, the source I2 was not detected in the VLA observations by \citet{DiGennaro2018ApJ...865...24D}. 
We see sharp edges along the length of both I1 \& I2, and there is no arc-like shape even at the junction of I1-I2.
Furthermore, both I1 and I2 have a similar brightness distribution at both frequencies (Fig. \ref{fig:400_700_8p5}). 
This makes them distinguishable from both the RN relic in the north and the surrounding radio halo (Fig. \ref{fig:I1_I2}). 
Source I1 has a width of $\sim 80$ kpc and extends $\sim 630$ kpc from the base of the RN relic towards relic R2, almost touching the east-west filament of the latter. Source I2 has a similar width, but a length of $\sim 400$ kpc. 

We estimated flux density values of I1 and I2 separately within the regions indicated in Fig. \ref{fig:Regions}. The flux densities of the source I1 are $S_{400} = 8.1 \pm 0.6$ and $S_{675} = 3.9 \pm 0.2$ mJy with $\alpha_{\rm I1} = -1.39 \pm 0.18$, and the source I2 are $S_{400} = 5.4 \pm 0.4$ and $S_{675} = 2.5 \pm 0.2$ mJy with $\alpha_{\rm I2} = -1.44 \pm 0.18$ (Table \ref{tab:int_spix}). 
The similarity in morphology and spectral indices of I1 and I2 may relate to their common origin.
However, despite their morphological differences, the spectral indices of the source I1 and I2 are also comparable to the typical spectral indices of halos and relics. 
\citet{Stroe2013A&A...555A.110S} found the spectral index of source I1 ($\alpha_\mathrm{0.608GHz}^\mathrm{1.382GHz} \sim -1.25$) to be similar to the typical radio halo spectral index and speculated it to be a flatter-spectrum part of the radio halo. 
On the other hand, because of their detection of an arc-like structure in front of the radio galaxy B, \citet{Hoang2017MNRAS.471.1107H} proposed it to be a bow shock generated with the interaction between source B and the RN relic downstream electrons. 
Also, the possibility of it being a filamentary object formed due to adiabatic shock compression \citep[][]{Enblin2001A&A...366...26E,Enblin2002MNRAS.331.1011E}{}{} is highly unlikely due to the lack of ultra-steep spectrum.
In contrast, our images reveal the object I1-I2 to have a narrow jet-like structure with a sharp corner, in addition to having similar brightness, width, length, and spectral index of the components I1 and I2. 
Although no compact radio source at the I1-I2 junction is detected, presence of a galaxy at cluster redshift (Fig. \ref{fig:I1_I2}) is reported by \citet{Dawson2015ApJ...805..143D}.
Therefore, we speculate this source to be a narrow-angle tail \citep[NAT,][]{Owen1976ApJ...205L...1O} radio galaxy.

\section{Southern Relics} \label{sec:SR}
The southern relics are detected at incredible detail at uGMRT sub-GHz frequencies (Fig. \ref{fig:Labels}, \ref{fig:400_700_8p5}). The bright filamentary `L' shaped component is labelled as RS1 and RS2. At lower brightness, RS3 and RS5 are also forming additional `arms' extending eastward, whereas RS4 extends towards the north. There is also presence of AGN activity J, from the source J1, embedded within the relic diffuse emission. Combining all of these components together as a single diffuse source, it stretches $\sim 2.5$ Mpc (LLS) in total from the east to west. This complex, diffuse source is discussed in more detail in the following sections.

\subsection{RS1 \& RS2 Relics}
The RS1-RS2 relic is one of the brightest components of the southern relic (Fig. \ref{fig:Labels}). Their peculiar `L' shape make them immediately distinguishable in that region.
From the junction of the `L', the relic RS1 extends towards the south-east, but a bifurcation can be seen at around $\sim 270$ kpc. The prominent RS1 filament continues up to $\sim 460$ kpc, but a fainter filament forks out from the bifurcation point and extends much further up to $\sim 540$ kpc in the east, with a width of $\sim 100$ kpc. 
On the other hand, the relic RS2 extends in the south-western direction from the `L' junction for $\sim 420$ kpc with a bend at the tip towards the south. Interestingly, this bend at the tip of the RS2 seems to align well with the overall southern edge of the south relic (see Fig. \ref{fig:400_700_8p5}). 

The brightness distribution along both relic RS1 and RS2 is inhomogeneous and patchy, with the brightest spot being the corner of the `L' shape at both 400 and 675 MHz. 
Furthermore, this bright spot seems to be a compact source that is clearly seen in high resolution images by \citet{DiGennaro2018ApJ...865...24D} i.e., at L-band in Fig. 1 and at S-band in Fig. 15.
An optical counterpart is indicated in Fig. \ref{fig:RS_arrows}, which was classified as cluster member by \citet{Dawson2015ApJ...805..143D}. 
Therefore, owing to its peculiar shape, brightness distribution, tail lengths and a compact source in the middle, the bright `L' shape in the RS1-RS2 system (Fig. \ref{fig:Labels}) seems more like a narrow-angle tail \citep[NAT,][]{Owen1976ApJ...205L...1O} radio galaxy embedded in that region.  Moreover, the lack of spectral steepening along the tails (typical of AGN jets) may be related to the recent re-energization by the shockwaves associated with the relics (Fig. \ref{fig:RS_spix_map_inj} \textit{left panel}).

\begin{figure*}
    \centering
    \includegraphics[width=2.1\columnwidth]{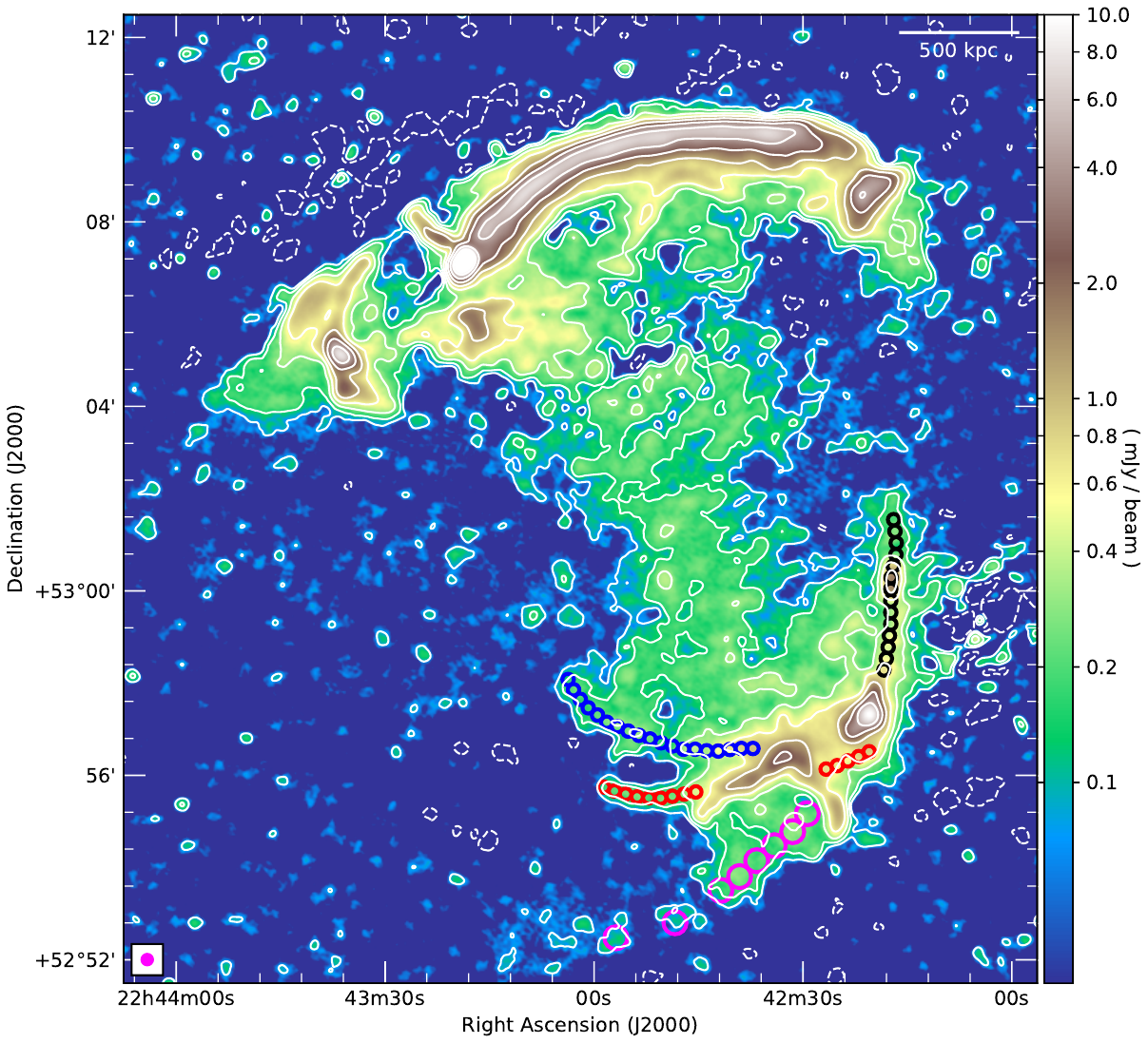}
    \caption{Low-resolution ($15\arcsec$) uGMRT image at 400 MHz overlaid with contours. The image properties are listed in Table \ref{tab:image_prop}, see label IM3. The contours are drawn at levels $[-1, 1, 2, 4, 8,...] \times 3\sigma_{\rm rms}$. Negative contours are shown with dashed line. The circular regions with beam width are used to extract spectral index profiles of the RS1 (red), RS3 (blue) and RS4 (black) relic. The RS5 relic profile is extracted with circles of width $30\arcsec$, indicated in magenta color. These profiles are presented in Fig. \ref{fig:RS1-RS5_spix_inj}.}
    \label{fig:400_bm15}
\end{figure*}

\begin{figure*}
    \centering
    \begin{tabular}{cc}
    \includegraphics[width=\columnwidth]{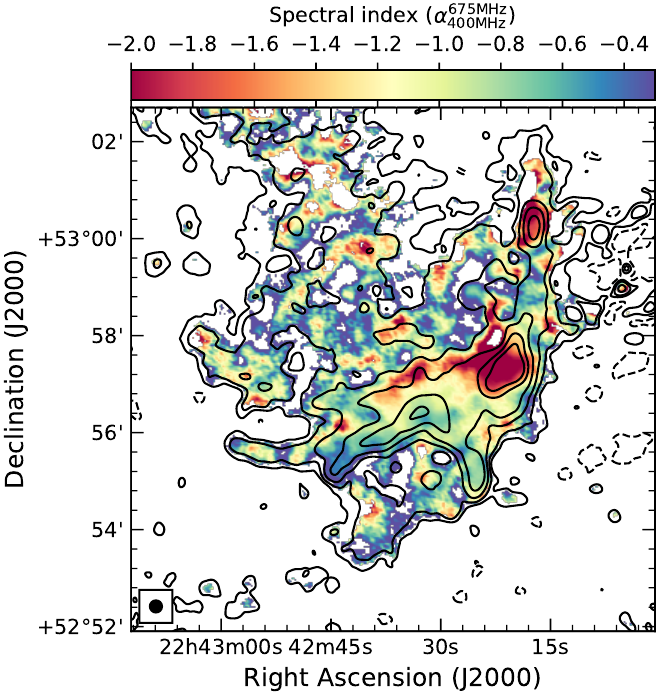}  & 
    \includegraphics[width=\columnwidth]{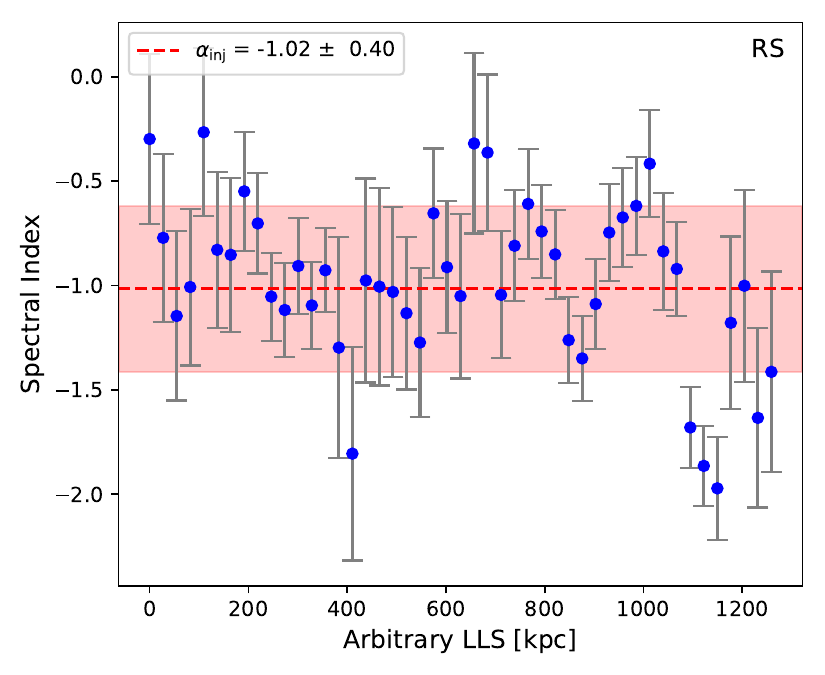}\\
    \end{tabular}
    \caption{(\textit{Left}): Low-resolution ($15\arcsec$) spectral index map of the southern relic RS between 400 and 675 MHz. The images used to make this map are IM3 and IM6 in Table \ref{tab:image_prop}. The contours overlaid on the map correspond to the 400 MHz image (IM3), and are drawn at levels $[-1, 1, 2, 4, 8,...] \times 3\sigma_{\rm rms}$. (\textit{Right}): Spectral index profile of the RS relic, combining the RS1, RS3 and RS4 regions shown in Fig. \ref{fig:400_bm15}.}
    \label{fig:RS_spix_map_inj}
\end{figure*}

From Fig. \ref{fig:RS_arrows}, we see that if the NAT radio galaxy is ignored as a separate diffuse source embedded in the region, the remaining low-brightness component of the RS1 can be traced, following the brightness variations, from the south of the source J to the far east end of the RS1, as indicated with red arrows. This total extent is $\sim 1.1$ Mpc, typical of radio relics \citep[e.g.,][]{vanWeeren2019SSRv..215...16V}{}{}. We suggest this to be the actual RS1 relic associated with the merging shock.
Furthermore, it is unlikely for the seed relativistic electrons from the source J to reach the far end of the RS1 relic, as suggested in the previous works, however, the eastern tail of the NAT radio galaxy seems a more plausible source in this regard.
Following the above reasoning, we derived the injection spectral index along the RS1 relic. For better S/N we used the $15\arcsec$ resolution image (see Fig. \ref{fig:400_bm15}) and estimated the injection spectral index to be $\alpha_{\rm inj} = -0.89 \pm 0.25$ (Fig. \ref{fig:RS1-RS5_spix_inj} \textit{top left}). This results in the associated shock Mach number being $\mathcal{M} = 2.5_{-0.4}^{+1.4}$, which in itself seems to be consistent with the shock Mach number of the whole RS relic derived in the previous studies (see Table \ref{tab:Mach_numbers}). 
On the other hand, since we considered RS2 to be one of the tails of the NAT radio galaxy instead of a radio relic \citep[][]{Stroe2013A&A...555A.110S,Hoang2017MNRAS.471.1107H,DiGennaro2018ApJ...865...24D}, we excluded deriving corresponding $\alpha_{\rm inj}$ and $\mathcal{M}$.

\begin{figure}
    \centering
    \includegraphics[width=\columnwidth]{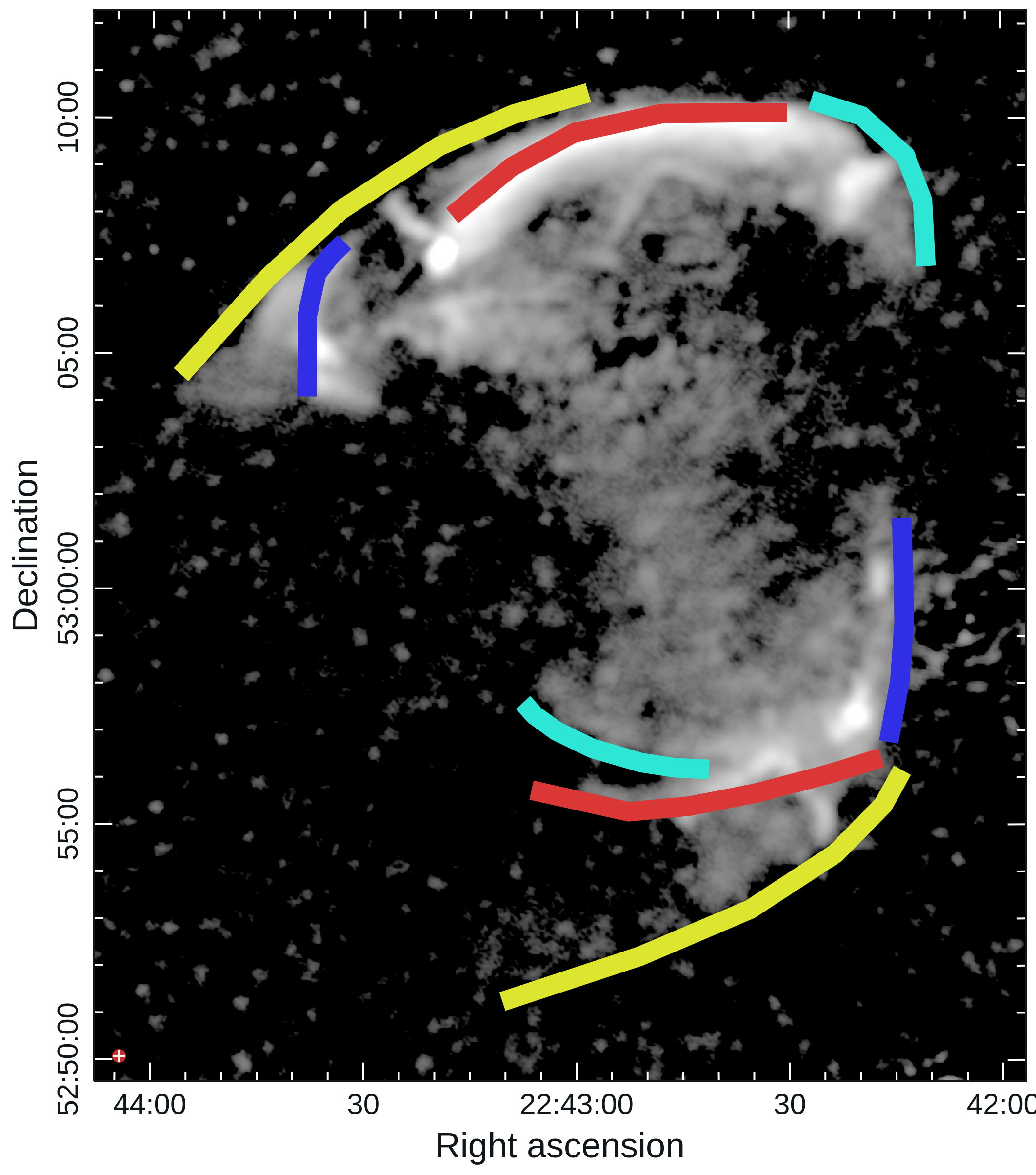}
    \caption{A schematic representation of the shock components possibly associated with the different relics in the `Sausage' cluster.}
    \label{fig:shock_arcs}
\end{figure}

\subsection{RS3 \& RS5 Relics}
RS3 and RS5 are two additional `arms' in the eastern half of the southern relic that extend far ahead of RS1 in the north-east and south-east direction, respectively (Fig. \ref{fig:400_bm15}). It is difficult to determine the western end of RS3, as it merges with the western part of RS1 near the RS1-RS2 NAT radio galaxy (Fig. \ref{fig:400_bm15}). Therefore, starting from the fork between RS1 and RS3, the length of the RS3 up to the eastern end (roughly indicated with magenta arrows in Fig. \ref{fig:RS_arrows}) is $\sim 960$ kpc, which is similar to the typical radio relic lengths \citep[e.g.,][]{vanWeeren2019SSRv..215...16V}{}{}. Although it is difficult to determine its width, as most of the northern edge merges with the radio halo, an estimation is made from the eastern tip to be $\sim 180$ kpc. Apart from the western part of RS3 that has similar brightness as RS1, the rest of the RS3 relic has fairly smooth and homogeneous low surface brightness distribution, as is evident from the contour maps in Fig. \ref{fig:400_bm15}, \ref{fig:700_bm15}. 

On the other hand, the RS5 relic is the faintest among the southern relics. The north-western part of the RS5 connected with the RS1 relic is the brighter half, whereas the continuing south-eastern half is only detected in patches at $\sim 3\sigma$ level in both frequencies. Similar to RS3, it is difficult to determine the true extent of the RS5 in the western end (Fig. \ref{fig:400_bm15}). However, measuring from the south of the source J (tracing the white arrows in Fig. \ref{fig:RS_arrows}) up to the south-eastern tip of the RS5 (see Fig. \ref{fig:400_bm15}) yields $\sim 1.8$ Mpc in length; wider in the northern half and narrower in the southern half. 
This is a re-confirmation of the faint southern extent of the RS5 relic that was first reported by \citet{Hoang2017MNRAS.471.1107H} with LOFAR observations, and were missing in the high-frequency VLA observations by \citet{DiGennaro2018ApJ...865...24D}.

Following a similar method as to RS1, we have estimated injection spectral index along the RS3 and RS5 relics (Fig. \ref{fig:400_bm15}). 
The derived injection spectral index and resulting Mach number of the RS3 relic are $\alpha_{\rm inj} = -0.97 \pm 0.32$ (Fig. \ref{fig:RS1-RS5_spix_inj} \textit{top right}) and $\mathcal{M} = 2.3_{-0.4}^{+1.5}$, respectively. 
Similar to the RS1 relic, we find the derived Mach number of the RS3 to be in agreement with values previously reported in the literature for the whole RS relic (see Table \ref{tab:Mach_numbers}). 

On the other hand, even with larger circular regions (width = $30\arcsec$) for better Signal-to-Noise (S/N), the derived injection spectral index of the RS5 relic is un-physically flat ($\alpha_{\rm inj} = -0.51 \pm 0.19$, see Fig. \ref{fig:RS1-RS5_spix_inj} \textit{bottom right}), even though the northern half of the RS5 relic has similar S/N ($\sim 6\sigma$) detection level at both frequencies (see Fig. \ref{fig:400_bm15}, \ref{fig:700_bm15}). 
The most possible reason for this is the missing flux at 400 MHz observations, as was previously mentioned for some of the faint components of the northern relics. 
More sensitive radio observations with a wider frequency range are necessary to determine the actual injection spectral index of the RS5 relic. 

As diffusion of seed electrons from the source J till the eastern end of RS3 seems unrealistic, contribution from other local sources with past AGN activity like [DJS2015] J224253.350+525706.073 (source A in Fig. \ref{fig:RS_arrows}) might be necessary. 
On the other hand, both the source J and the NAT RS1-RS2 can supply seed electrons to the bright western half of the RS5 relic, however, the fainter Mpc scale south-eastern extent needs other explanation as no AGN source is identified in the region.

\subsection{RS4 Relic}
The RS4 relic extends from the region of source J almost directly north (yellow arrows in Fig. \ref{fig:RS_arrows}) to $\sim 1.2$ Mpc. 
The southern half of the relic between source N (Fig. \ref{fig:RS_arrows}) and J is brighter than the northern half at both frequencies (Fig. \ref{fig:400_bm15}, \ref{fig:700_bm15}). 
Surprisingly, this source N was detected as a bright spot in most of the previous reports \citep[e.g.,][]{Stroe2013A&A...555A.110S,Hoang2017MNRAS.471.1107H}, except for \citet{DiGennaro2018ApJ...865...24D}. 
Furthermore, it also has very steep spectrum ($\alpha \lesssim -1.7$) like source J within the relic RS4 (Fig. \ref{fig:RS_spix_map_inj} \textit{left panel}). 
We speculate that this source may be similar to the source J i.e., an AGN tail possibly associated with the adjacent compact source on the east N1 (Fig. \ref{fig:RS_arrows}). Moreover, the presence of strong magnetic fields in the AGN lobe \citep[e.g.,][]{Hardcastle1998MNRAS.294..615H}{}{} can explain its higher brightness than the surroundings at low frequencies \citep[][and this work]{Stroe2013A&A...555A.110S,Hoang2017MNRAS.471.1107H}{}{} and its faintness or total absence at higher frequencies \citep[][]{Stroe2013A&A...555A.110S,DiGennaro2018ApJ...865...24D}{}{}, resulting in the observed extremely steep spectrum.
Avoiding the region with source J contamination (Fig. \ref{fig:400_bm15}), we estimated injection spectral index and corresponding Mach number of the RS4 relic to be $\alpha_{\rm inj} = -1.17 \pm 0.56$ (Fig. \ref{fig:RS1-RS5_spix_inj} \textit{bottom left}) and $\mathcal{M} = 2.0_{-0.4}^{+2.3}$, respectively. The large uncertainty in $\alpha_{\rm inj}$ is the result of large variation in the spectral index along the relic length, associated with the low-brightness region (too flat) and the source N (too steep). 
Better observations are needed for more accurate spectral measurement. 
Finally, if the source N is confirmed to be an AGN tail (at high resolution), it can also be the source of fossil plasma for the RS4 relic, 
similar to the previous argument by \citet{DiGennaro2018ApJ...865...24D} in the case of source J for the whole RS relic.
Furthermore, the brighter southern half can receive contribution from both J and N, whereas only source N provides for the fainter northern half of the relic.

\subsection{Source J \& J1} \label{subsec:J}
Source J is the brightest region in the southern relic at both frequencies. While \citet{Stroe2013A&A...555A.110S} speculated it to be a radio phoenix embedded within the southern relic, \citet{DiGennaro2018ApJ...865...24D} revealed it to be connected with an AGN labelled as J1. At both 400 and 675 MHz, we detect the same J1 AGN as well as its connection with the source J (see Fig. \ref{fig:RS_arrows}), re-confirming the previous findings. 
In the spectral index map (Fig. \ref{fig:RS_spix_map_inj} left) we find this source to have the steepest spectrum ($\alpha \lesssim -2$)within the southern relic. 
Although source J1 is subtracted in most of the images shown in this work, the narrow jet has a slightly flatter spectrum near the AGN J1, progressively steepening and again flattening towards the shock front, confirming the spectral variation trend shown in Fig. 13 of \citet{DiGennaro2018ApJ...865...24D}.

\subsection{South Relic RS} \label{subsec:RS}
Since we have considered the southern relic to be composed of individual components as discussed above, rather than one single giant radio relic, it is not possible to directly compare our results with previous works.
Therefore, for the purpose of comparison of the RS relic with the previous works, we estimated the integrated flux densities of the whole RS relic (see Fig. \ref{fig:Regions}) to be $S_{400} = 170.8 \pm 12.0$ and $96.7 \pm 5.8$ mJy. The integrated spectral index comes out to be $-1.09 \pm 0.18$ (Table \ref{tab:int_spix}), which is in agreement with result by \citet{DiGennaro2018ApJ...865...24D}. 
Since a significant part of the southern relic is much fainter compared to the northern relics, the flux density measurements are performed on $15\arcsec$ images (Fig. \ref{fig:400_bm15}, \ref{fig:700_bm15}).

Although there is some evidence of spectral steepening towards the cluster center (Fig. \ref{fig:RS_spix_map_inj} \textit{left}), the quality of the map is not robust enough to comment on this with high confidence. On the other hand, a weak gradient in spectral index in the east-west direction can be seen from the $\alpha_{\rm inj}$ of RS1 and RS4, as was previously reported by \citet{DiGennaro2018ApJ...865...24D}. 
We have also derived a representative injection spectral index of the RS relic by combining the $\alpha_{\rm inj}$ profiles of the RS1, RS3 and RS4 together in Fig. \ref{fig:RS_spix_map_inj} (\textit{right panel}). The resultant injection spectral index came out to be $\alpha_{\rm inj} = -1.02 \pm 0.40$ with a corresponding Mach number of $\mathcal{M} = 2.2_{-0.4}^{+2.1}$. Apart from the large positive uncertainty, the derived Mach number is in agreement with the previous works \citep[][]{Akamatsu2015A&A...582A..87A,Hoang2017MNRAS.471.1107H,DiGennaro2018ApJ...865...24D}{}{}.  

We see that $\alpha_{\rm inj}$ derived from different regions of the RS relic in this work (Fig. \ref{fig:400_bm15}), in \citet{DiGennaro2018ApJ...865...24D} (Fig. 11), and in \citet{Hoang2017MNRAS.471.1107H} (Fig. 9) results in similar Mach numbers (see Table \ref{tab:Mach_numbers}). 
This may have the implication that (i) either the whole RS relic, with all its complex structures, is related to a single merging shock wave of the derived Mach number as the previous works suggested, or (ii) the estimated Mach number is an averaged value of different individual concurrent shocks with similar Mach numbers associated with the merging event.
In Fig. \ref{fig:shock_arcs}, we present a schematic representation of these shock components, and speculate in favor of the latter scenario, in that the observed complex filamentary structure of the southern relic is the result of multiple concurrent shock waves, associated with the merging event, energizing the ICM electrons. 
Moreover, the seed electrons were likely supplied from local sources like the NAT RS1-RS2, A, J and N (Fig. \ref{fig:RS_arrows}). 
In fact, the presence of a shock at the R1 relic position \citep{Hoang2017MNRAS.471.1107H} supports this scenario. More sensitive X-ray observations are necessary in confirming the presence of shocks associated with other individual relic components. 
We would like to note that these proposed shock structures across the cluster are based purely on the symmetry of the observed complex morphology of the northern and southern relics. Dedicated detailed simulations and/or improved observations are needed to validate these claims, which is beyond the scope of this work.

\section{Summary and Conclusions} \label{sec:conclude}
In this work, we used uGMRT band-3 and band-4 data to observe the morphological structures of diffuse radio objects in the `Sausage' cluster in detail. We re-confirmed several morphological and spectral features that were previously reported in the literature, as well as discovered some new features. A brief summary of the results are presented below.

\noindent (i) Northern relics
\begin{itemize}
    \item[--] We observe variation in spectral index profiles across the relic width, from the east of the RN relic to the west. This may indicate a decrease in the downstream cooling rate in the east-west direction. Also, the Mach number ($\mathcal{M_\mathrm{RN}} = 2.9_{-0.2}^{+0.3}$) measured in this work is in agreement with the previous results. 
    \item[--] We re-confirm the presence of $\sim 930$ kpc R5 relic in the north of RN.
    \item[--] Relic R1 arc shows sub-structure in the spectral profile, which may indicate the presence of finer filaments. Presence of spectral gradient is observed from the relic filament towards the downstream region. 
    \item[--] The $\sim 840$ kpc R4 relic has two distinct components, one bright and filamentary, and the other faint diffuse non-filamentary. Presence of a double-strand structure similar to the `Toothbrush' relic is detected. The Mach number of the relic is found to be $\sim 2.7$. 
    \item[--] On the western side of the RN, the relic R3 has a bright component with a sharp brightness edge towards the south-west, then further extending with a faint diffuse component in the same direction. Spectral index gradient towards the cluster center is observed in both these components. We further notice similarity in patchy spectral index distribution with the adjacent RN5 relic component. Finally, we estimate the associated shock Mach number of the R3 to be $\sim 2.8$. 
    \item[--] We report the detection of source I2, perpendicular to source I1 (previously known as source I) in front of the radio galaxy B. Unlike the arc-like structure of a possible bow shock previously reported by \citet{Hoang2017MNRAS.471.1107H}, we find a lot of evidence of it being a NAT radio galaxy. 
\end{itemize}

\noindent (ii) Southern relics RS1-RS2
\begin{itemize}
    \item[--] We categorize the bright `L' shaped structure to be a NAT radio galaxy, and a source of seed electrons for RS1 relic.
    \item[--] Tracing the brightness variations in the high-resolution images, we suggest the actual RS1 relic stretches up to the south of the source J (LLS $\sim 1.1$ Mpc). 
    \item[--] Relic width towards the east end is $\sim 100$ kpc, making it one of the narrower relics till reported \citep[e.g.,][]{vanWeeren2010Sci...330..347V}{}{}.
    \item[--] The derived Mach number of the RS1 relic is $\sim 2.5$.
\end{itemize} 

\noindent (iii) Southern relics RS3 \& RS5
\begin{itemize}
    \item[--] We re-confirm the detection of relic RS3 at 400 and 675 MHz. The projected length of the RS3 relic is $\sim 960$ kpc with a Mach number of $\sim 2.3$.
    \item[--] We measure the length of RS5 to $\sim 1.8$ Mpc, making it one of the larger radio relics discovered so far \citep[e.g.,][]{Giovannini1991A&A...252..528G,vanWeeren2010Sci...330..347V,vanWeeren2016ApJ...818..204V,Rajpurohit2021A&A...654A..41R}{}{}. This is a re-confirmation of the fainter south extension in the RS5 that was first detected by \citet{Hoang2017MNRAS.471.1107H} but failed to be confirmed by \citet{DiGennaro2018ApJ...865...24D} with their high-frequency observations.
    \item[--] We suggest both the RS1-RS2 NAT and J to be the source of seed electrons for the RS5 relic.
\end{itemize}

\noindent (iv) Southern relic RS4
\begin{itemize}
    \item[--] Extremely steep spectrum source N, embedded within RS4, seems similar to the source J associated with AGN activity. We speculate it to be the possible source of fossil plasma in the RS4 relic.
    \item[--] The estimated Mach number of the $\sim 1.2$ Mpc scale RS4 relic is found to be $\sim 2.0$. 
\end{itemize}

\noindent (v) Southern relic RS
\begin{itemize}
    \item[--] Considering the observed features discussed above, we suggest the RS relic to be a union of the component relics, each associated with a different shock, unlike the previous assumption of a single shock wave responsible for the whole complex southern relics system. Deeper X-ray observations are necessary for further investigation. 
    \item[--] Based on the morphological symmetry observed between northern and southern relics, we suggest a schematic shock structure associated with the merger event, which needs to be verified with simulation and better observations.  
    \item[--] We re-confirm the presence of the AGN J1 and its filamentary connection with source J. We further suggest that along with J, local AGN sources like the NAT, A and N also plays critical role in providing necessary seed relativistic electrons for the component relic formation. 
\end{itemize}

In this work, we showed that with sensitive high-resolution radio observations, a lot of new features can be uncovered that were previously missed. However, the observational data used in this work suffer a number of limitations, which lead to larger uncertainty in the spectral analysis. Future better observations of this kind over a wide frequency range with good \textit{uv-}coverage is needed for more robust spectral analysis and better understanding of the ongoing astrophysical processes. Furthermore, simulations of the merger event that led to the formation of the observed complex diffuse radio structures, composed of radio halo and multiple relics, will provide key insights into their formation mechanism that are still severely lacking. Finally, very sensitive X-ray observations are needed to explore the presence/absence of shocks at the individual relic positions and their corresponding X-ray Mach numbers to compare with the radio derived results.  

\begin{acknowledgments}
We would like to thank Rhodes University for providing the necessary computing facilities for data analysis. We thank the staff of GMRT, who made these observations possible. GMRT is run by the National Centre for Radio Astrophysics of the Tata Institute of Fundamental Research. 
RR and OS's research is supported by the South African Research Chairs Initiative of the Department of Science and Technology and the National Research Foundation (Grant id. 81737).
MR and HYKY acknowledge support from the National Science and Technology Council (NSTC) of Taiwan (NSTC 112-2628-M-007-003-MY3). HYKY acknowledges support from the Yushan Scholar Program of the Ministry of Education (MoE) of Taiwan.

\end{acknowledgments}

%

\vspace{5mm}
\facilities{uGMRT}


\software{\texttt{CASA} \citep{McMullin2007ASPC..376..127M,CASA2022PASP..134k4501C}, 
\texttt{CARTA} \citep{angus_comrie_2018_3377984},
\texttt{SPAM} \citep{Intema2009A&A...501.1185I,Intema2017A&A...598A..78I},
\texttt{WSCLEAN} \citep{offringa-wsclean-2014,Offringa2017MNRAS.471..301O},
\texttt{Astropy} \citep{AstropyCollaboration2013A&A...558A..33A,AstropyCollaboration2018AJ....156..123A,Astropy2022ApJ...935..167A}, \texttt{APLpy} \citep{aplpy}, \texttt{Matplotlib} \citep{matplotlib}, \texttt{DS9} \citep{DS92003ASPC..295..489J}}



\appendix

\begin{figure*}[]
    \centering
    \begin{tabular}{cc}
    \includegraphics[width=0.48\columnwidth]{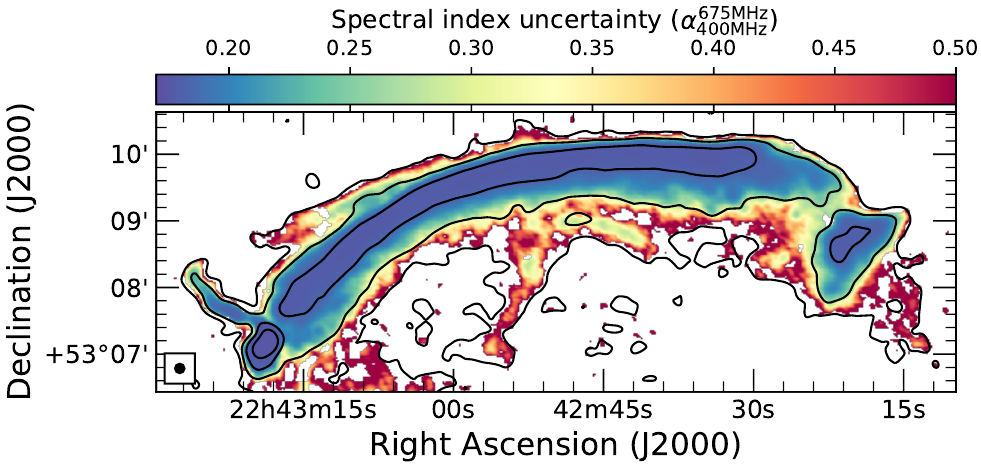} & 
    \includegraphics[width=0.48\columnwidth]{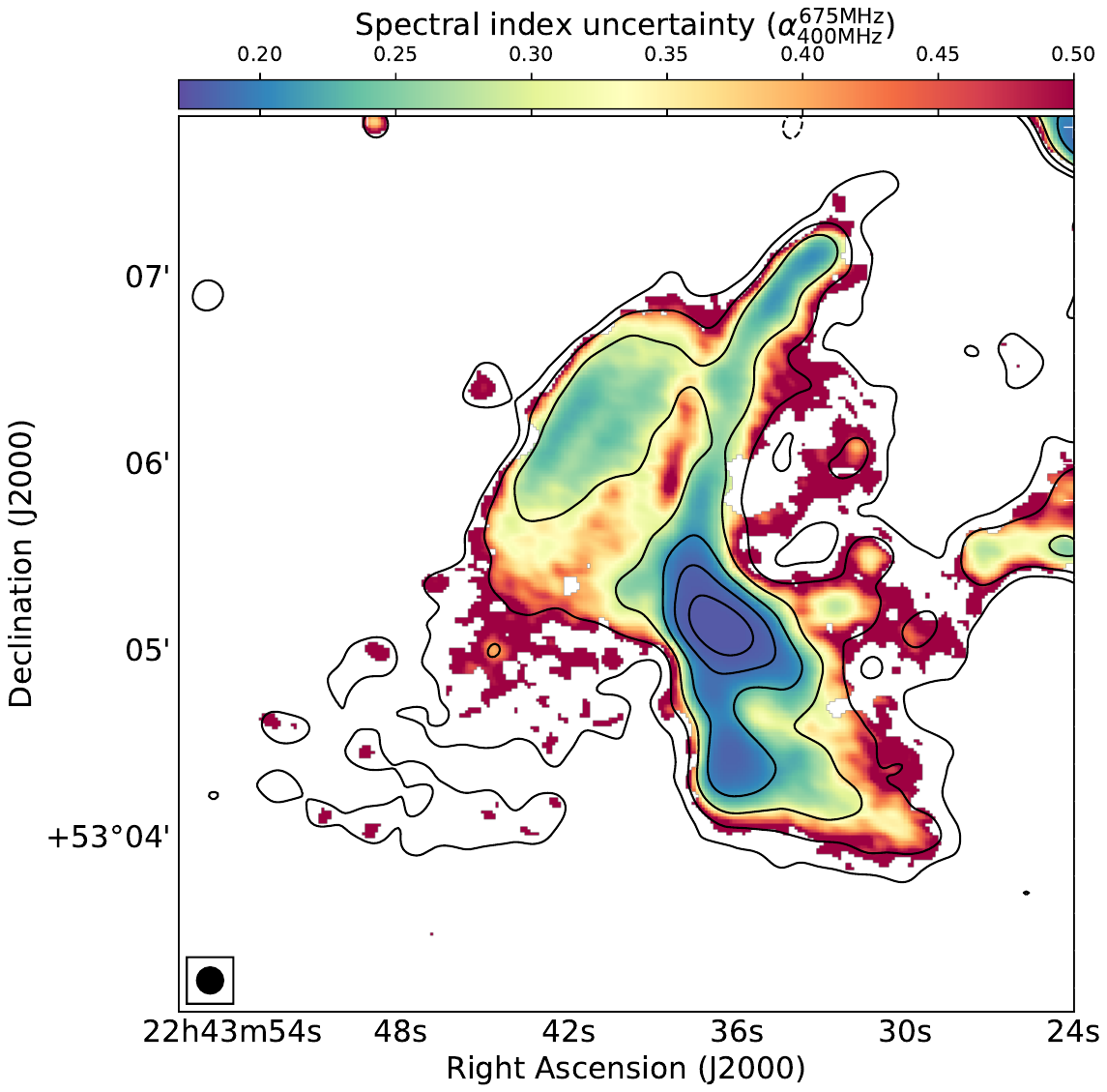} \\
    \includegraphics[width=0.48\columnwidth]{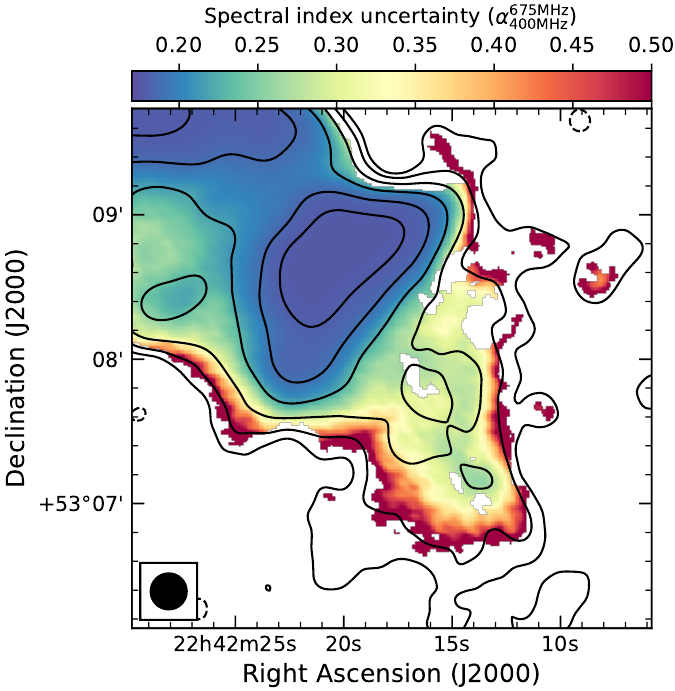} &
    \includegraphics[width=0.48\columnwidth]{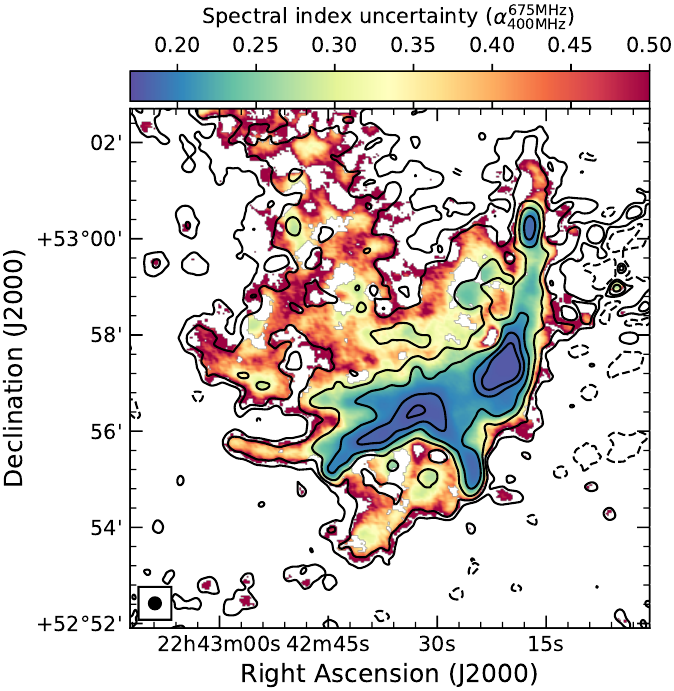}
    \end{tabular}
    \caption{Spectral index uncertainty maps corresponding to Fig. \ref{fig:RN_spix_map} in (\textit{top left}) panel, Fig. \ref{fig:R1_R4_linear_spix} in (\textit{top right}) panel, Fig. \ref{fig:R3_maps} in (\textit{bottom left}) panel, and Fig. \ref{fig:RS_spix_map_inj} in (\textit{bottom right}) panel.}
    \label{fig:spix_err_maps}
\end{figure*}

\begin{figure*}
    \centering
    \includegraphics[width=\columnwidth]{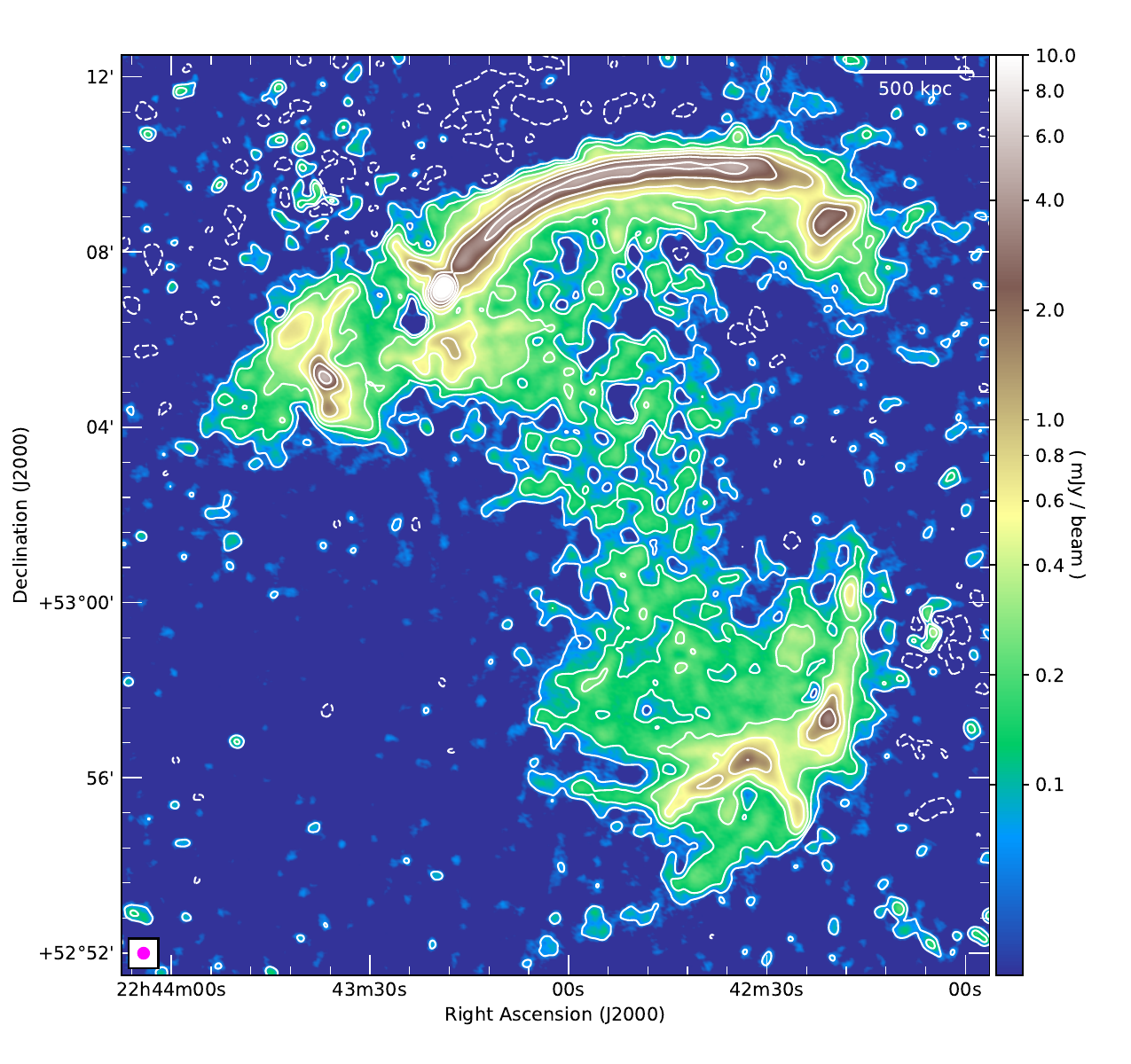}
    \caption{Low-resolution ($15\arcsec$) uGMRT image at 675 MHz overlaid with contours. The image properties are listed in Table \ref{tab:image_prop}, see label IM6. The contours are drawn at levels $[-1, 1, 2, 4, 8,...] \times 3\sigma_{\rm rms}$. Negative contours are shown with dashed line.}
    \label{fig:700_bm15}
\end{figure*}

\begin{figure*}[]
    \centering
    \begin{tabular}{cc}
    \includegraphics[width=0.48\columnwidth]{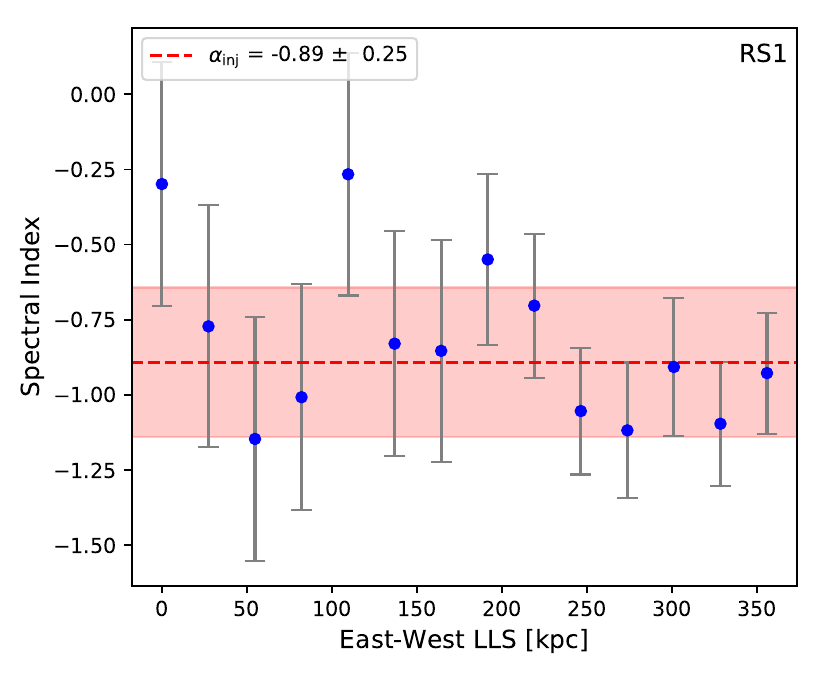} & 
    \includegraphics[width=0.48\columnwidth]{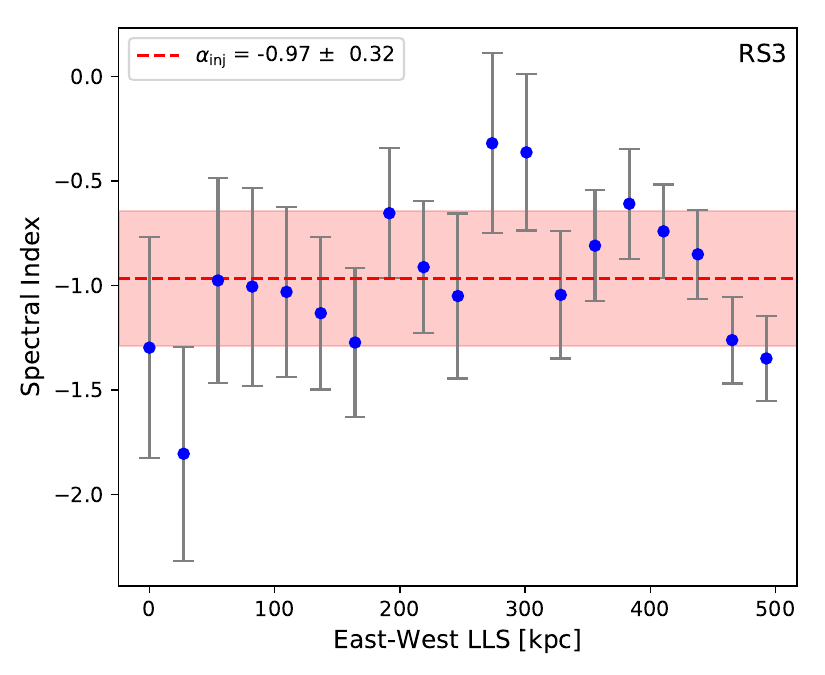} \\
    \includegraphics[width=0.48\columnwidth]{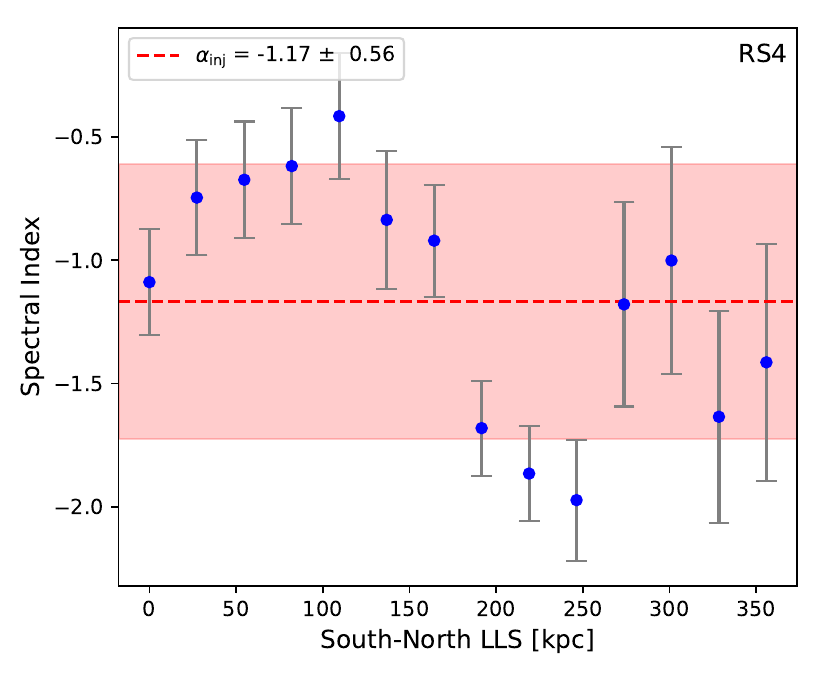} & 
    \includegraphics[width=0.48\columnwidth]{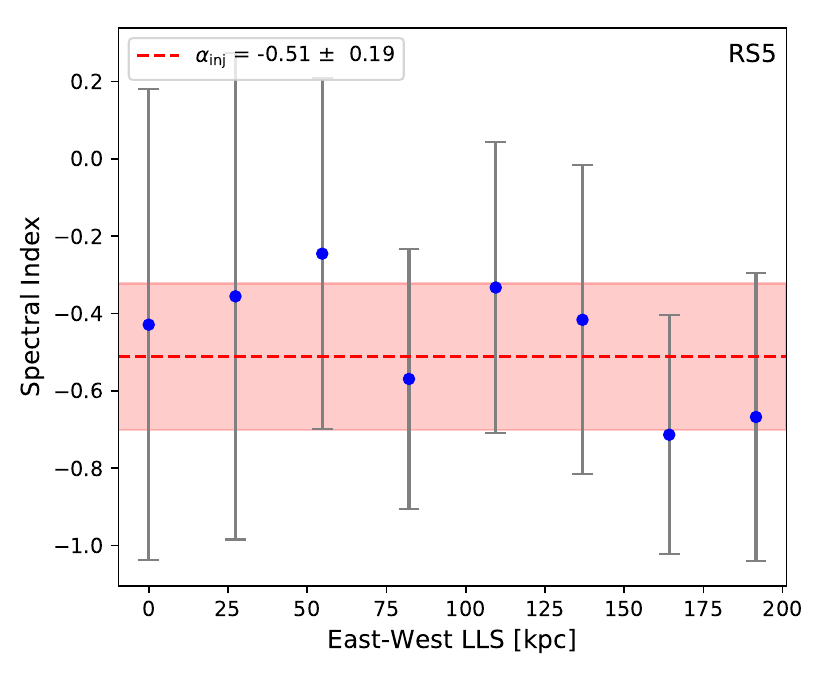}
    \end{tabular}
    \caption{Spectral index profiles of the RS1 (\textit{top left}), RS3 (\textit{top right}), RS4 (\textit{bottom left}) and RS5 (\textit{bottom right}) relic, corresponding to the regions shown in Fig. \ref{fig:400_bm15}.}
    \label{fig:RS1-RS5_spix_inj}
\end{figure*}


\bibliography{myBib}{}
\bibliographystyle{aasjournal}



\end{document}